\documentclass[prb,showpacs,twocolumn,aps]{revtex4}
\usepackage[dvipdfm]{graphicx}
\usepackage[dvipdfm]{color}
\usepackage{dcolumn}
\usepackage{amsmath}
\usepackage{amssymb}

\newcommand{\lco}{$\rm La_{2}CuO_4$}
\newcommand{\lsco}{$\rm La_{2-{\textit x}}Sr_\textit{x}CuO_4$}

\newcommand{\lsfco}{$\rm La_{1.85}Sr_{0.15}CuO_4$}

\newcommand{\leco}{$\rm La_{1.8}Eu_{0.2}CuO_4$}
\newcommand{\lescoxy}{$\rm La_{2- \textit{x} - \textit{y}}Eu_\textit{y}Sr_\textit{x}CuO_4$}
\newcommand{\lesco}{$\rm La_{1.8- \textit{x}}Eu_{0.2}Sr_\textit{x}CuO_4$}
\newcommand{\lesaco}{$\rm La_{1.782}Eu_{0.2}Sr_{0.018}CuO_4$}
\newcommand{\leszco}{$\rm La_{1.78}Eu_{0.2}Sr_{0.02}CuO_4$}
\newcommand{\lesfco}{$\rm La_{1.65}Eu_{0.2}Sr_{0.15}CuO_4$}

\newcommand{\lscoh}{$\bf La_{2-x}Sr_xCuO_4$}
\newcommand{\lecoh}{$\bf La_{1.8}Eu_{0.2}CuO_4$}
\newcommand{\lescoh}{$\bf La_{1.8-x}Eu_{0.2}Sr_xCuO_4$}

\newcommand{\lecohn}{$\bf La_{1.8}Eu_{0.2}CuO_4$}

\newcommand{\lnco}{$\rm La_{1.7}Nd_{0.3}CuO_4$}
\newcommand{\keimer}{$\rm La_{1.65}Nd_{0.35}CuO_4$}
\newcommand{\lndco}{$\rm La_{2- \textit{y}}Nd_\textit{y}CuO_4$}
\newcommand{\lsmco}{$\rm La_{2- \textit{y}}Sm_\textit{y}CuO_4$}
\newcommand{\lsmzco}{$\rm La_{1.8}Sm_{0.2}CuO_4$}

\newcommand{\scoc}{$\rm Sr_{2}CuO_2Cl_2$}

\newcommand{\ybco}{$\rm YBa_{2}Cu_3O_{6+\delta}$}

\newcommand{\sus}{susceptibility}
\newcommand{\schi}{$\chi$}

\newcommand{\DM}{Dzyaloshinsky--Moriya}

\newcommand{\pla}{$\rm CuO_2$}
\newcommand{\oct}{$\rm CuO_6$}

\newcommand{\HAF}{$\chi_{2DHAF}$}

\begin{document}
\title{Dzyaloshinsky--Moriya Spin Canting in the LTT Phase of $\bf La_{2-x-y}Eu_ySr_xCuO_4$}
\author{M.~H\"ucker$^{1}$, V.~Kataev$^{2}$, J.~Pommer$^{2}$, U.~Ammerahl$^{3}$, A.~Revcolevschi$^{3}$, J.~M.~Tranquada$^{1}$, and B.~B\"uchner$^{4}$}
\affiliation{$^1$Physics Department, Brookhaven National Laboratory, Upton, New York 11973}
\affiliation{$^2$II. Physikalisches Institut, Universit\"at zu K\"oln,
D-50937 K\"oln, Germany}
\affiliation{$^3$Labaratoire de Chimie des Solides, Universit\'e
Paris-Sud, F-91405 Orsay C\'edex, France}
\affiliation{$^4$Institut f\"ur Festk\"orper- und Werkstofforschung
Dresden, D-01171 Dresden, Germany}
\date{\today}
\begin{abstract}
The Cu spin magnetism in \lescoxy\ ($x\leq 0.17$; $y\leq0.2$) has been
studied by means of magnetization measurements in fields up to 14~Tesla.
Our results clearly show that in the antiferromagnetic phase \DM\ (DM)
superexchange causes Cu spin canting not only in the LTO phase but also in
the structural low-temperature phases LTLO and LTT. In \leco\ the canted
DM-moment is about 50\% larger than in pure \lco\  which we attribute to
the larger octahedral tilt angle in the Eu-doped compound. We also find
clear evidence that the size of the canted DM-moment does not change
significantly at the structural transition at $T_{LT}$ from LTO to LTLO
and LTT. The most important change induced by the transition is a
significant reduction of the magnetic coupling between the $\rm CuO_2$
planes. As a consequence, the spin-flip transition of the canted Cu spins
which is observed in the LTO phase for magnetic field perpendicular to the
$\rm CuO_2$ planes disappears in the LTT phase. The shape of the
magnetization curves changes from the well known spin-flip type to a
weak-ferromagnet type. However, no spontaneous weak ferromagnetism is
observed even at very low temperatures, which seems to indicate that the
interlayer decoupling in our samples is not perfect. Nonetheless, a small
fraction ($\lesssim 15$\%) of the DM-moments can be remanently magnetized
throughout the entire antiferromagnetically ordered LTT/LTLO phase, i.e.
for $T<T_{LT}$ and $x<0.02$. It appears that the remanent DM-moment is
perpendicular to the $\rm CuO_2$ planes. For magnetic field parallel to
the $\rm CuO_2$ planes we find that the critical field of the spin-flop
transition decreases in the LTLO phase, which might indicate a competition
between different in-plane anisotropies. To study the Cu spin magnetism in
\lescoxy , a careful analysis of the Van Vleck paramagnetism of the $\rm
Eu^{3+}$ ions was performed.
\end{abstract}
\maketitle

\section{Introduction}
\label{introduction}

In rare-earth (RE) doped \lsco\ the structural transition from the
low-temperature orthorhombic phase (LTO) to the low-temperature tetragonal
phase (LTT) has attracted a lot of attention because it causes a
suppression of the superconducting ground state in favor of an
incommensurate antiferromagnetic (AF) charge and spin stripe
order.~\cite{Axe89,Crawford91,Buechner94c,Tranquada95a,Wagener97a,Klauss00a}
On the other hand, it is found that the low-temperature (LT) transition
has a considerable impact on the AF order in insulating ($x=0$) and
lightly-Sr-doped compounds ($x\lesssim 0.02$). Though several
experimental~\cite{Shamoto92,Crawford93,Keimer93,Suh98b,Suh99b,Kataev99b,Tsukada03a}
and
theoretical~\cite{Bonesteel93b,Shekhtman93,Viertio94a,Koshibae94a,Yildirim95,Stein96}
studies have focused on the AF regime, important details of the spin
structure in the LTT phase are still a matter of debate. In particular,
the question in dispute is whether or not the magnetic groundstate in the
LTT phase has a canted spin structure, as is the case in the LTO phase.

In the LTO phase of \lco\ the N\'eel temperature ($T_N\simeq 325$~K) as
well as the spin structure are determined by a combination of small
anisotropic contributions to the superexchange, $J \simeq 135$~meV, and a
weak interlayer coupling, $J_\perp \sim 10^{-5} \times
J$.~\cite{Thio88,Kastner88,Lyons88a,Aeppli89,Hayden91,Kastner98} As is
shown in Fig.~\ref{spinstructure1}(a), the resulting magnetic ground state
is a collinear spin structure with Cu spins almost parallel to the
$b$-axis (spacegroup $Bmab$) but slightly canted out-of-plane ($\parallel
c$) by $\sim 0.2^\circ$.~\cite{Thio88,Kastner88,Thio94} Since for a
particular layer all spins cant into the same direction each layer carries
a weak ferromagnetic (WF) moment $\parallel c$.~\cite{Thio88} Spin canting
originates from the \DM\ (DM) antisymmetric anisotropic superexchange,
$J_{DM}\sim 10^{-2} \times J$, which is directly coupled to the tilting of
the \oct\ octahedra by an angle of the order of $5^\circ$ at low
temperatures.~\cite{Moriya60,Dzyaloshinsky58,Coffey91,Bonesteel93b}

In the LTO phase the gain in DM exchange energy, and hence the canting
angle, is maximum for Cu spins perpendicular to the octahedral tilt axis
$\parallel [100]$ [cf. Fig.~\ref{spinstructure1}(b)]. In RE-doped \lco\ at
the LT-transition the octahedral tilt axis rotates azimuthally by an angle
$\alpha$, whereby axes in adjacent layers rotate in opposite
directions.~\cite{Axe89} The question that arises is, whether the spins
follow the rotation of the tilt axes to maintain a maximum gain in DM
exchange energy [black spins in Fig.~\ref{spinstructure1}(b)], or whether,
as was predicted theoretically,~\cite{Stein96} they rotate in the opposite
direction, resulting in a spin structure without spin canting [white spins
in Fig.~\ref{spinstructure1}(b)].

In this paper we present a detailed study of the Cu spin magnetism in the
LTO and LTT phases of \lescoxy\ in magnetic fields up to 14~Tesla. Our
results unambiguously show that DM spin canting exists in the LTT phase.
Furthermore, the size of the DM moments does not change at the structural
transition LTO$\leftrightarrow$LTT within the error of the experiment.
These findings support a spin structure in the LTT phase where, due to the
DM superexchange, the Cu spins remain perpendicular to the octahedral tilt
axis [black spins in Fig.~\ref{spinstructure1}(b)]. Compared to pure \lco
, the DM moment in \leco\ is about 50\% larger in both the LTO and LTT
phases, which we attribute to a larger octahedral tilt angle. The most
import difference between LTO and LTT is a significant reduction of the
interlayer coupling in the LTT phase which macroscopically results in the
disappearance of the spin-flip transition and in a considerable increase
of the \sus\ and magnetization at low magnetic fields. To be able to
directly compare the Cu spin magnetism in Eu-doped samples with pure \lsco
, we have put considerable effort into the careful analysis and
subtraction of the Van Vleck magnetism of the $\rm Eu^{3+}$ ions.

The paper is organized as follows: In Sec.~\ref{background} we provide
some more detailed background knowledge. In Sec.~\ref{experimental} we
briefly discuss the investigated stoichiometries, experimental details and
relevant aspects of the crystal structure. In Sec.~\ref{VanVleck} we
determine the Van Vleck magnetism of the $\rm Eu^{3+}$ ions, which was
necessary to uncover the Cu spin magnetism. In Sec.~\ref{doping} we
present a general overview of the Cu spin magnetism as a function of Sr
and Eu doping. Most of these measurements were performed at 1~Tesla. In
the subsequent Sec.~\ref{pure}, we focus on the \DM\ spin canting and the
in-plane gap in pure \leco . We present the high-field magnetization up to
14~Tesla for a polycrystal and a single crystal. In Sec.~\ref{canting} we
analyze the \DM\ spin canting in this pure compound. Resulting phase
diagrams are discussed in Sec.~\ref{diagrams}. In Sec.~\ref{lightly} we
show results for lightly-Sr-doped \leco\ polycrystals ($0 \leq x \leq
0.02$) and directly compare them with pure \lsco . A discussion is given
in Sec.~\ref{discussion}.

\section{Background}
\label{background}

\begin{figure}[t]
\center{\includegraphics[width=0.78\columnwidth,angle=0,clip]{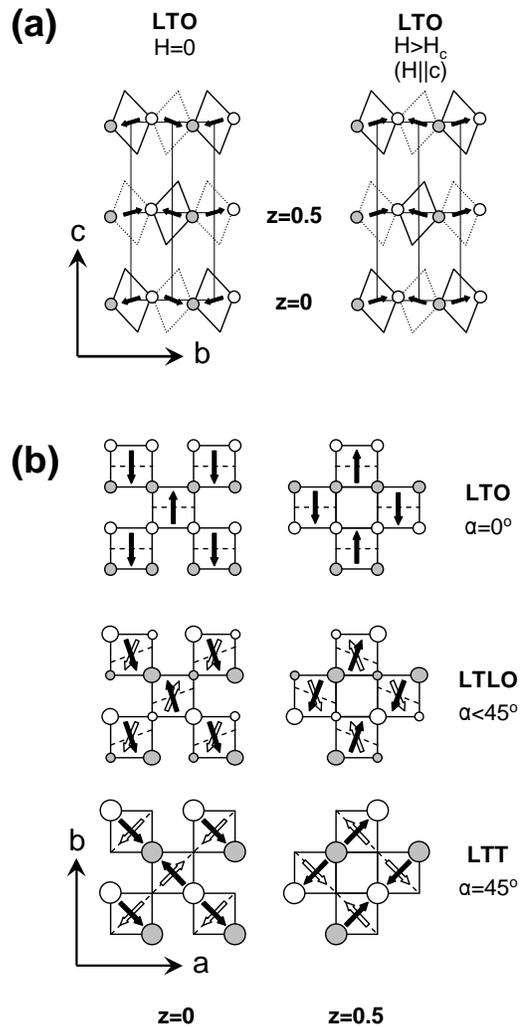}}
\caption[]{Spin structure (a) in LTO phase with antiferromagnetically
($H=0$) and weak ferromagnetically ($H>H_c$ parallel $c$-axis) coupled
$\rm CuO_2$ planes. Octahedral tilts and spin canting exaggerated. (b)
spin structure in two adjacent $\rm CuO_2$ planes in the LTO, LTLO and LTT
phase. Grey (white) oxygen atoms are displaced below (above) $\rm CuO_2$
plane. Size of circles grows with displacement. Dashed lines indicate the
octahedral tilt axis. Black arrows: Cu spins follow azimuthal rotation of
tilt axis, i.e. spins are always perpendicular to the tilt axis. DM spin
canting in LTT phase maximum. White arrows: spins rotate in opposite
direction. In the LTT phase spins are parallel to tilt axis. There is no
DM spin canting. Figure (b) after Vierti\"o and
Bonesteel.~\cite{Viertio94a}} \label{spinstructure1}
\end{figure}

It is well known that the azimuthal rotation of the octahedral tilt axis
below the LT-transition can take values in the range $0^\circ < \alpha <
45^\circ$. In contrast, the octahedral tilt angle itself changes only very
little at the transition.~\cite{BradenDiss} The LTT phase with spacegroup
$P4_2/ncm$ is stabilized when $\alpha=45^\circ$, while for
$\alpha<45^\circ$ the low-temperature-less-orthorhombic phase (LTLO) with
spacegroup $Pccn$ is formed, which is an intermediate phase between LTO
and LTT [cf. Fig.~\ref{spinstructure1}(b)]. In particular in
lightly-Sr-doped samples containing excess-oxygen $\delta$, the LTLO phase
with $\alpha$ much smaller than $45^\circ$ is formed. By reducing the
excess-oxygen concentration, $\alpha$ increases, and for $\delta
\rightarrow 0$ the LTLO structure approaches LTT. In particular, for $x=0$
and $\delta$ just slightly larger than zero it was shown that the
LT-transition is better described by a sequence of transformations: a
discontinuous transition LTO$\rightarrow$LTLO, followed by a continuous
transition LTLO$\rightarrow$LTT.~\cite{Crawford97a}

In the LTT phase the octahedral tilt axes are oriented alternately along
[110] and [1\={1}0] directions in adjacent \pla\ planes (cf.
Fig.~\ref{spinstructure1}). As a result, the Cu-O-Cu bonds of a particular
layer are buckled only in one crystallographic direction and the DM
superexchange is active only for spin components pointing along the
direction of buckling. Hence, for the LTT phase two principle magnetic
ground states are possible: one with and another without DM spin canting,
depending on whether the spins stay perpendicular (black spins) or
parallel (white spins) to the octahedral tilt axis. Similarly, in the LTLO
phase the spins are assumed to be either perpendicular to the tilt axis or
rotated by an angle of $2\alpha$ relative to the perpendicular spin
direction. Which of these spin structures (white/black spins) represents
the ground state in the LTLO and LTT phase is the subject of this paper.

Early magnetization measurements on \lndco\ and \lsmco\ revealed a
remanent moment $M_{REM}$ in the LTT phase.~\cite{Shamoto92,Crawford93}
This result, in combination with neutron diffraction data, led the authors
of Refs.~\onlinecite{Shamoto92} and~\onlinecite{Crawford93} to the
conclusion that DM spin canting exists in the LTT phase. It happens,
however, that in \lndco\ $M_{REM}$ increases with increasing Nd content
and shows a Curie-type temperature dependence. As a result, it has been
questioned whether the weak ferromagnetism emerges from a Nd-Cu
interaction.~\cite{Stein96} Neutron diffraction experiments by Keimer et
al. on \lndco\ indicate that at the LT-transition the Cu spins follow the
azimuthal rotation of the octahedral tilt axis, i.e., the spins remain
perpendicular to the tilt axis (black spins in
Fig.~\ref{spinstructure1}).~\cite{Keimer93} Furthermore, the authors argue
that in the LTT phase the tetragonal crystal symmetry, in combination with
the non-collinear spin structure, should lead to a frustrated interlayer
exchange. Magnetization measurements of \lesco\ polycrystals with $0 \leq
x\leq 0.02$ in Ref.~\onlinecite{Kataev99b} are in general agreement with
the conclusions of the neutron diffraction study. Since non-magnetic $\rm
Eu^{3+}$ was used instead of magnetic $\rm RE^{3+}$, these measurements
rule out the possibility that the observed LTT ground state with spin
canting is the result of RE-Cu interactions. In contrast to these
findings, recent magnetization data on a \leco\ single crystal with
$T_N=265$~K were interpreted in favor of spin structures given by the
white spins in Fig.~\ref{spinstructure1}, which means that DM spin canting
is reduced in the LTLO phase and disappears in the LTT
phase.~\cite{Tsukada03a}

Several theoretical papers have addressed the problem of DM spin canting
in the LTT phase, and there is general agreement that the magnetic ground
state depends on a delicate balance between various contributions to the
in-plane and out-of-plane anisotropies, as well as the interlayer
coupling.~\cite{Shekhtman92,Shekhtman93,Bonesteel93b,
Viertio94a,Koshibae94a,Yildirim94b,Yildirim95,Stein96} In particular, it
was claimed that the in-plane spin direction depends on a competition
between the DM interaction and a symmetric superexchange anisotropy, which
in the case of RE-doped \lco\ is supposed to favor a LTT phase without
spin canting (white arrows in
Fig.~\ref{spinstructure1}).~\cite{Shekhtman92,Shekhtman93,Bonesteel93b,
Viertio94a} Although the most recent calculations have confirmed this
result,~\cite{Stein96} there are serious discrepancies with experiment
which remain to be understood. According to
Shekhtman~et~al.,~\cite{Shekhtman93} in \lco\ the symmetric anisotropy
would cause in-plane and out-of-plane spin-wave gaps of equal size, i.e.,
an Ising-type anisotropy. In contrast, neutron scattering experiments show
that in \lco\ and several other layered cuprates the out-of-plane gap is
larger than the in-plane gap and always of the order of
5~meV.~\cite{Keimer93,Greven95,Matsuda90,Yildirim94b} Shekhtman et
al.,~\cite{Shekhtman93,Yildirim94b} as well as other groups, suggest that
this discrepancy follows from a contribution to the out-of-plane gap from
direct exchange, which seems not to depend significantly on the ligand
structure perpendicular to the \pla\ plane. In particular, isostructural
\scoc\ with flat \pla\ planes has the same out-of-plane gap as \lco\ but a
much smaller in-plane gap. This seems to show that octahedral tilting
mainly tunes the in-plane-gap and, therefore, puts a very low limit on a
possible contribution of the Ising-type symmetric anisotropy to both
spin-wave gaps. From this perspective, one can say that the idea of the
symmetric superexchange anisotropy dominating the antisymmetric DM
superexchange is at least experimentally not well settled in the two
dimensional (2D) spin $S=1/2$ cuprate Heisenberg antiferromagnets.

\section{Experiment}
\label{experimental}
The polycrystalline samples of \lescoxy\ used in this study were prepared
by a standard solid state reaction~\cite{Breuer93}. Three series of
samples with the following Sr and Eu concentrations have been
investigated: one with fixed Eu content ($y=0.2$) and various Sr dopings
($0\leq x\leq0.17$), and two others with fixed Sr concentrations
($x=0.017$; 0.08) and various Eu dopings ($0\leq x \leq 0.2$). For
comparison, we will present results on a \lnco\ and several \lsco\
polycrystals. The \lesco\ single crystals with $x=0$ and 0.15 were grown
using the traveling-solvent floating-zone method. The magnetization
measurements were performed with two different magnetometers, a Faraday
balance (4-330~K; 0-1~Tesla) as well as a vibrating-sample magnetometer
(4-290~K; 0-14~Tesla). Sample masses have varied between 0.3 and 0.7~g.
Some of the samples were also investigated with $\rm
\mu$SR,~\cite{Klauss99b,Klauss00a,Klauss02a,Klauss03a}
NQR,~\cite{Suh98b,Suh99b,Curro00a,Simovic03a,Simovic03b}
ESR~\cite{Gadolinium,Kataev97a,Kataev98a,Kataev99b} and
x-rays~\cite{JKlugeDipl}. Very limited normal state magnetization data on
the $\rm Eu_{0.2}$-doped samples were previously published in
Refs.~\onlinecite{Huecker97,Kataev99b,Simovic03a}. Throughout this paper,
magnetization curves $M(H)$ are presented in units of $\rm \mu_B$/Cu. To
convert the data into units of $\rm G\cdot emu /mol$ a factor of
$1/1.79055\times 10^{-4}$ has to be applied.

As was mentioned in Sec.~\ref{background}, interstitial oxygen in samples
with low Sr content has a considerable effect on the crystal structure. To
remove the excess-oxygen, samples were annealed under reducing conditions.
Most magnetization data on polycrystals were obtained after anneals for
3~d at 625$^\circ$C in flowing $\rm N_2$ gas. Two \lesco\ specimens with
$x<0.02$, studied by x-ray diffraction, have shown a discontinuous
LTO$\leftrightarrow$LTLO transition at $T_{LT}\sim 135K$ and a continuous
transition LTLO$\leftrightarrow$LTT at around 60~K, which is well below
$T_{LT}$.~\cite{JKlugeDipl} Later on, the magnetization of some
polycrystals was remeasured after they were annealed for 1/2\,h at
800$^\circ$C in vacuum ($\lesssim 10^{-4}$~mbar), which results in lower
values of $\delta$. These samples exhibit 10-20~K higher values of $T_N$
and very sharp structural and magnetic transitions. In the case of the
polycrystal with $x=0$, $T_{N}$ increased from 285~K to 316~K.

The smaller the RE ionic radius, the harder the structural ground state is
pushed towards the LTT phase. For Nd$^{3+}$- and Sm$^{3+}$-doped \lsco\
this has been shown by Crawford et al. in Ref.~\onlinecite{Crawford97a}.
Since Eu$^{3+}$ is even smaller, it induces the LTO$\leftrightarrow$LTLO
transition already at a lower level of doping, and at increasing Eu
content, the LTT phase is stabilized up to higher temperatures. Moreover,
with respect to the average ionic radius at the La site, $\rm Eu_{0.2}$ is
comparable to $\rm Sm_{0.23}$ and $\rm Nd_{0.36}$.~\cite{Shannon76a} Since
in our Eu doped $x=0$ sample annealed at 800$^\circ$C the N\'eel
temperature is comparable to the values in Ref.~\onlinecite{Crawford97a}
(i.e. comparable $\delta$), we assume that the $T$ range below $T_{LT}$ in
which the LTLO phase exists is even smaller than in \lsmzco .
Corresponding low temperature data for \lesco\ will therefore be discussed
in terms of the LTT phase.

Similar arguments apply for our \leco\ single crystal, which was studied
first after annealing for 3~d at 625$^\circ$C in $\rm N_2$ and a second
time after annealing for 2~h at 800$^\circ$C in vacuum. It seems, though,
that the oxygen content is slightly higher and less homogenous than in the
\leco\ polycrystal, possibly because of the longer diffusion path. Our
\sus\ data indeed shows that after annealing at 625$^\circ$C, below
$T_{LT}$ the crystal stays in the LTLO phase down to 4~K. After the second
annealing at 800$^\circ$C it approaches the LTT phase at $T\simeq 100K$.

In our experience with \lesco , excess-oxygen affects the LT-transition up
to a Sr content as high as $x=0.08$, which is considerably higher than in
pure \lsco . For $x>0.08$ it was confirmed by x-ray diffraction
experiments that in as-prepared samples the LT-transition is of the
LTO$\leftrightarrow$LTT type.~\cite{JKlugeDipl} Where it is relevant, we
will indicate the samples' annealing history in the text.

\begin{figure}[b]
\center{\includegraphics[width=0.9\columnwidth,angle=0,clip]{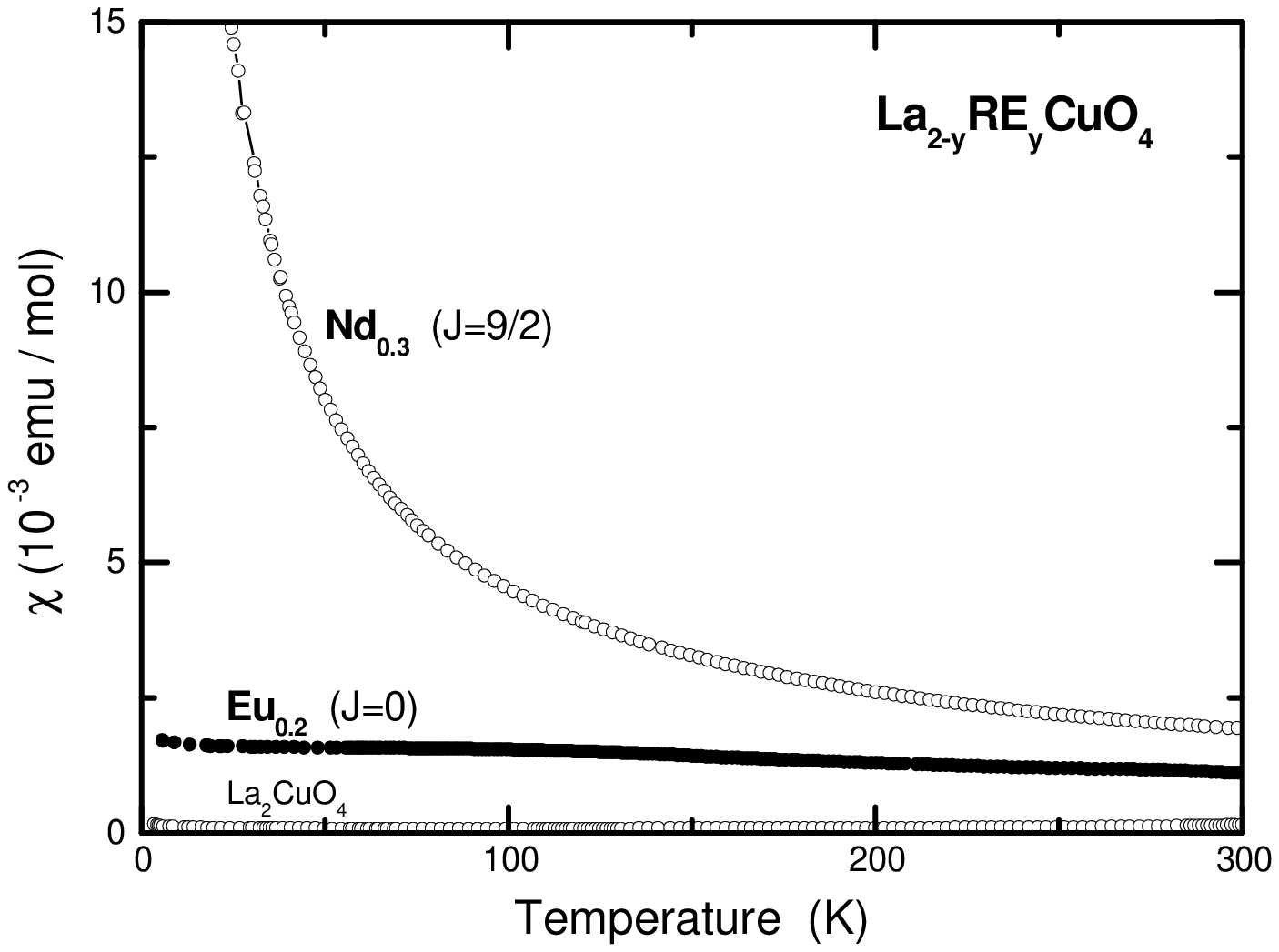}}
\caption[]{Static magnetic \sus\ ($H=1$~Tesla) of pure and RE-doped \lco\
with $\rm RE_y = Nd_{0.3}$ and $\rm Eu_{0.2}$.} \label{SEvgl}
\end{figure}
\section{Anisotropic E\lowercase{u}$\rm ^{3+}$ Van Vleck magnetism}
\label{VanVleck}
An important reason for using europium is that $\rm Eu^{3+}$ ions have a
non-magnetic ground state $\rm J=0$, whereas other RE ions such as $\rm
Nd^{3+}$ ($\rm J=9/2$) and $\rm Sm^{3+}$ ($\rm J=5/2$) have a magnetic
ground state. Hence, by choosing Eu we avoid a possible impact of the
RE--Cu interaction on the Cu spin magnetism. $\rm Eu^{3+}$ ions exhibit
Van~Vleck-paramagnetism which is constant at low temperatures and only
moderately changes at higher temperatures, as one can see in
Fig.~\ref{SEvgl}.~\cite{VanVleck} In contrast, $\rm Nd^{3+}$ ions exhibit
a strongly temperature-dependent Curie-type \sus\ which makes it
impossible to analyze the much weaker Cu-spin magnetism, as the comparison
with \lco\ demonstrates. We mention that Sm-doped samples show a \sus\
that is even smaller than for Eu-doped samples.~\cite{Crawford97a}

\begin{figure}[b]
\begin{center}
\includegraphics[angle =0,width=0.48\textwidth,clip]{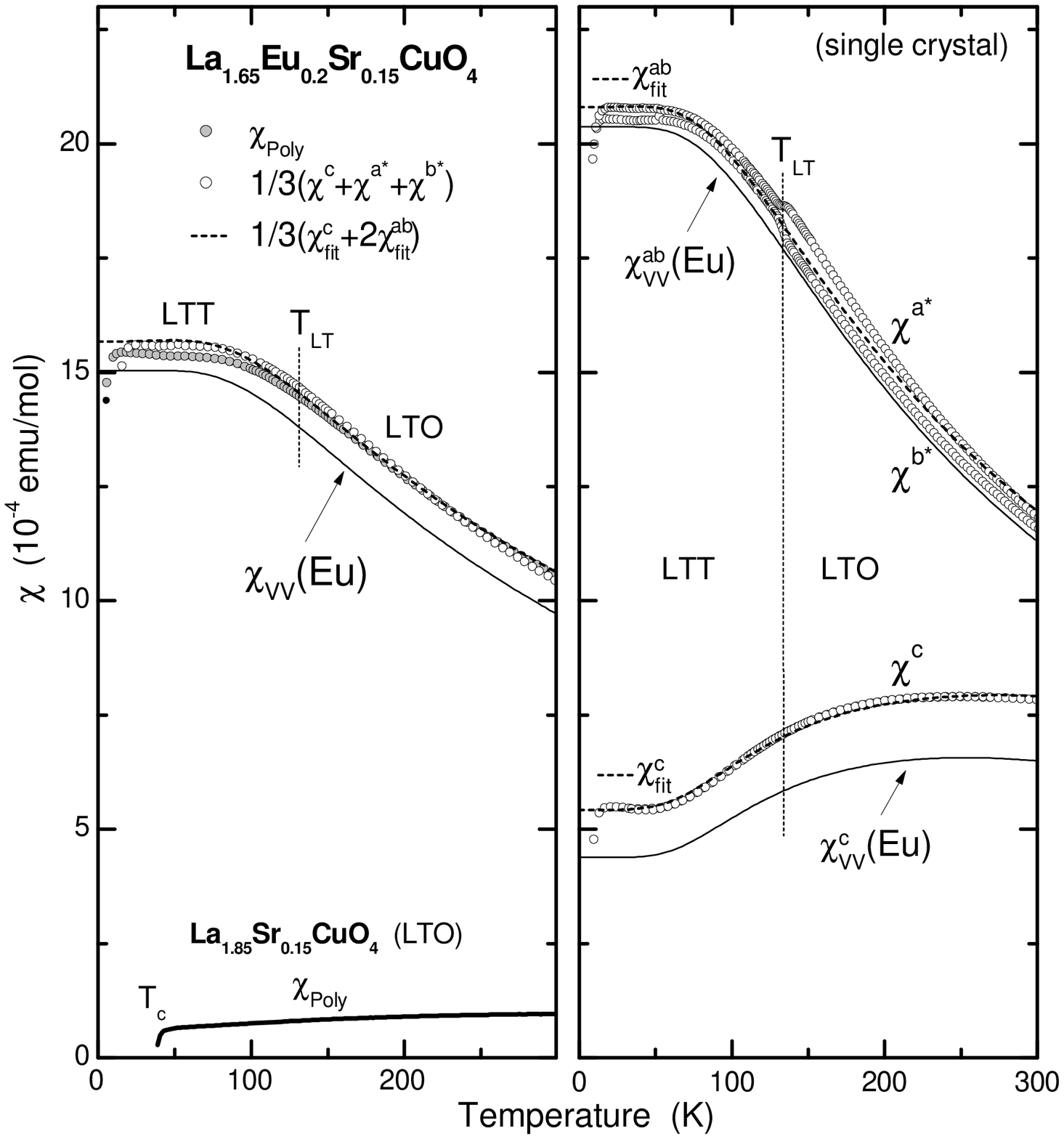}
\end{center}
\vspace{-0.5cm} \caption[]{Static magnetic \sus\ ($H=1$~Tesla) of \lesfco
. Left: comparison of polycrystal data with averaged single crystal data
as well as the \sus\ of a \lsfco\ polycrystal. $T_{LT}$ indicates the
LTO$\leftrightarrow$LTT transition, $T_c$ the SC transition in the pure
compound. Right: \sus\ of the single crystal for all three directions.}
\label{U2_J2zwei_pap3}
\end{figure}

In the following we briefly describe the analysis of the $\rm
Eu^{3+}$-Van~Vleck-paramagnetism, $\chi_{VV}$(Eu), which we will then
subtract from the total \sus\ to uncover the Cu-spin magnetism. The
analysis of $\chi_{VV}$(Eu) was performed for $x=0.15$, since at this Sr
content the magnetic contribution of the $\rm CuO_2$ planes is less
complicated than in the AF phase at low $x$. In Fig.~\ref{U2_J2zwei_pap3},
we show the \sus\ of a polycrystal (left) and for all three directions of
a single crystal (right). In the left hand plot we show also the averaged
\sus\ of the single crystal, which almost perfectly agrees with the
polycrystal data. Compared to pure \lsfco\ the \sus\ of the $\rm
Eu_{0.2}$-doped polycrystal is about one order of magnitude larger, due to
the dominant contribution of $\chi_{VV}$(Eu). At low temperatures the
\sus\ is constant, and for $T \gtrsim 70$~K it decreases monotonically. In
the case of the single crystal, $\chi_{VV}$(Eu) shows a strong
crystal-field anisotropy, which we have already described in
Ref.~\onlinecite{Simovic03a}. The \sus\ for magnetic field $H\parallel c$
is much smaller than for $H\parallel ab$ and increases with increasing
temperature. The structural transition LTO$\leftrightarrow$LTT occurs at
$T_{LT}\simeq 135$~K and is only visible for $H\parallel ab$-plane. It
shows up due to a small $ab$-anisotropy in the LTO phase that is
macroscopically eliminated in the LTT phase.~\cite{anisotropy} As was
explained in Ref.~\onlinecite{Simovic03a}, the crystal direction which
shows the negative (positive) step at $T_{LT}$ with decreasing temperature
predominantly contains the $a$-axis ($b$-axis) and is therefore called
$\chi^{a^*}$ ($\chi^{b^*}$) in terms of spacegroup $Bmab$. For $H\parallel
c$, no signature at $T_{LT}$ is observed, as was also checked for crystals
with $x=0.04$, 0.08, 0.12, and 0.2.~\cite{HueckerDiss} In the case of the
polycrystal, the LT-transition causes only a very small anomaly, as will
be discussed later (see also Ref.~\onlinecite{Huecker97}). The dotted
lines in Fig.~\ref{U2_J2zwei_pap3} are fits according to the following
expression:
$$ \chi^{ab,c}_{fit}=  \chi^{ab,c}_{LSCO} + \chi^{ab,c}_{VV}(Eu,y^*)$$
where $\chi_{LSCO}$ is the \sus\ of pure \lsco\ and $y^*$ is the
Eu-fraction as determined from the fit. For $H\parallel ab$ we have fitted
the average $ab$-plane \sus\ $(\chi^{a^*} + \chi^{b^*})/2$. To account for
$\chi_{LSCO}$, we have approximated the single crystal data in
Ref.~\onlinecite{Terasaki92}. The Van Vleck magnetism (solid lines) was
calculated using a similar approach as in Ref.~\onlinecite{Tovar89}, which
we have improved so that $\chi_{VV}$(Eu) is correctly described at higher
temperatures which was a major problem in Ref.~\onlinecite{Tovar89} and
\onlinecite{Rettori96}. Details of the calculation will be presented
elsewhere. The fits provide an almost perfect description of the data
which suggests that we have accurate expressions for $\chi_{VV}$(Eu).
Note, that this is an important ingredient for our analysis of the Cu-spin
magnetism in the AF phase for $x\leq 0.02$, where the determination of
$\chi_{VV}$(Eu) is considerably complicated by magnetic transitions of the
Cu spins at $T_N$ and $T_{LT}$.
\begin{figure}[b]
\center{\includegraphics[width=0.95\columnwidth,angle=0,clip]{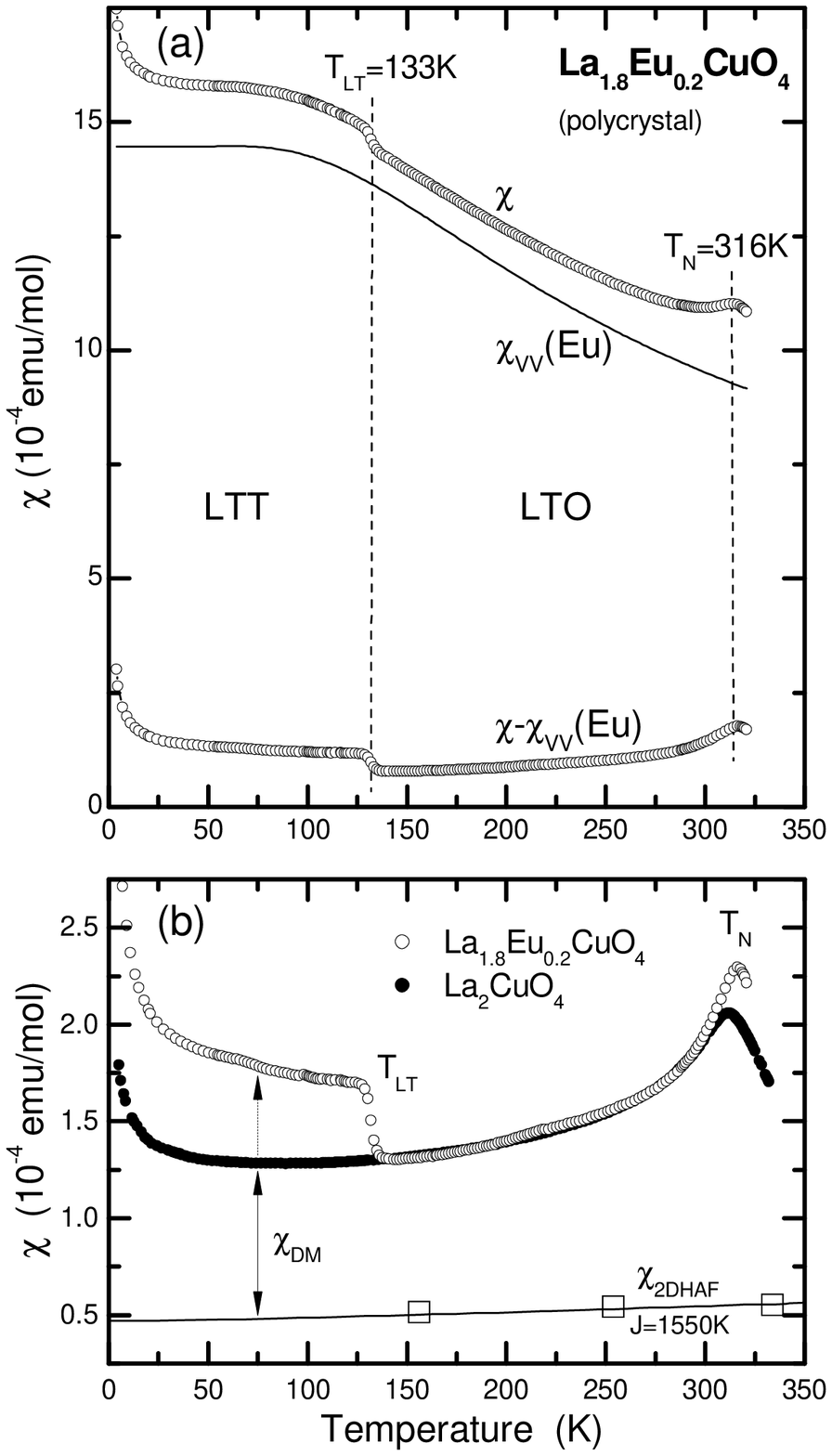}}
\caption[]{Static \sus\ ($H=1$~Tesla) of a $\rm La_{1.8}Eu_{0.2}CuO_4$
polycrystal. (a) Total \sus\ $\chi$, calculated Eu Van Vleck \sus\
$\chi_{VV}$(Eu), as well as difference $\chi - \chi_{VV}$(Eu). The dotted
lines indicate the antiferromagnetic transition at $T_N$ and the
structural transition at $T_{LT}$. (b) Comparison of $\chi -
\chi_{VV}$(Eu) with $\chi$ of pure \lco . Data in (b) corrected for
$\chi_{dia}=-0.99 \times 10^{-4}$~emu/mol and $\chi_{VV}{\rm (Cu)}=0.43
\times 10^{-4}$~emu/mol. ($\square$) \sus\ $\chi_{2DHAF}$ of a S1/2-2D-HAF
for $J=1550$~K (after Monte Carlo data in Ref.~\onlinecite{Okabe88b}).}
\label{eu_ah88_pap_combi}
\end{figure}

As an example for a sample in the AF phase, we show in
Fig.~\ref{eu_ah88_pap_combi}~(a) the \sus\ $\chi$ of a \leco\ polycrystal
($x=0$). Two transitions are visible: the N\'eel order transition at
$T_N=316$~K and the LTO$\leftrightarrow$LTT transition at $T_{LT}=133$~K.
The solid line is the Van Vleck fit that we have subtracted to obtain
$\chi - \chi_{VV}$(Eu). In Fig.~\ref{eu_ah88_pap_combi}~(b) we compare
$\chi - \chi_{VV}$(Eu) with the \sus\ of \lco . The Eu fraction $y^*$ was
varied so that the curves coincide in the LTO phase at 140~K, which
obviously causes
\begin{figure}[t]
\begin{center}
\includegraphics[angle =0,width=0.3\textwidth,clip]{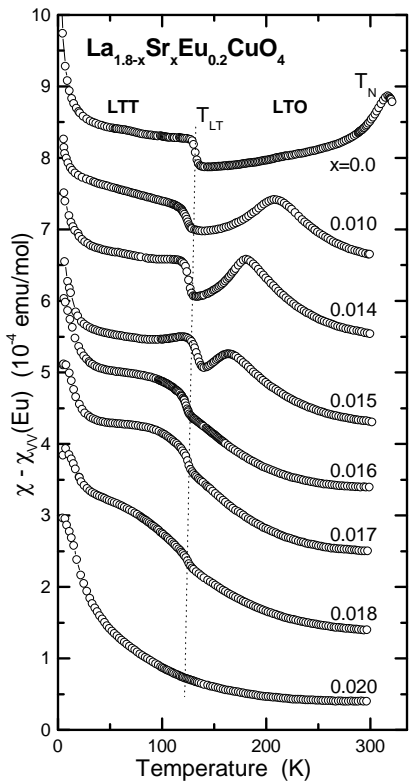}
\end{center} \vspace{-0.5cm} \caption[]{Static \sus\
($H = 1$~Tesla) of \lesco\ polycrystals for different Sr concentrations
$0\leq x \leq 0.02$ after subtraction of $\chi_{VV}$(Eu). Except for $x=0$
(800~$^\circ$C vac.) all samples were annealed at 625~$^\circ$C in $\rm
N_2$ gas. Curves are shifted for clarity by $1\times 10^{-4}$~emu/mol.}
\label{c_smallx2_pap}
\end{figure}
coincidence in a broad temperature range of the LTO phase. This procedure
yields a Eu fraction of $y^*=0.195(5)$ which matches the nominal
concentration of $y=0.2$ within the experimental error. At the
LT-transition $\chi - \chi_{VV}$(Eu) shows a step-like increase which
leads to deviations of the order of $0.5\times 10^{-4}$~emu/mol from the
\sus\ of \lco . At low temperatures both curves show a small Curie-type
increase which indicates a small number of free spins. As this
contribution is of same magnitude in both samples, we can exclude that Eu
doping leads to a significant increase of free moments or magnetic
impurities such as $\rm Eu^{2+}$ ions. In Fig.~\ref{eu_ah88_pap_combi}~(b)
we also show the \sus\ \HAF\ of a
spin $S=1/2$ 2D Heisenberg antiferromagnet for
$J=1550$~K.~\cite{Okabe88b,Johnston97a} To compare \HAF\ to the measured
curves, the latter ones have been corrected for the core diamagnetism
$\chi_{dia}$ and the Van Vleck magnetism $\chi_{VV}$(Cu) of the $\rm
Cu^{2+}$ ions.~\cite{Allgeier93,Landolt} In the LTO phase, the difference
$\chi_{DM}$ between the data and \HAF\ follows from the DM spin canting.
Without spin canting, the \sus\ in both the LTO and LT phase would be very
close to \HAF , as is the case at high temperatures in the high
temperature tetragonal (HTT) phase with flat $\rm CuO_2$ planes. In
contrast, in the LT phase the contribution of $\chi_{DM}$ even increases,
which clearly indicates that DM spin canting cannot be absent for
$T<T_{LT}$.
\begin{figure}[t]
\begin{center}
\includegraphics[angle =0,width=0.48\textwidth,clip]{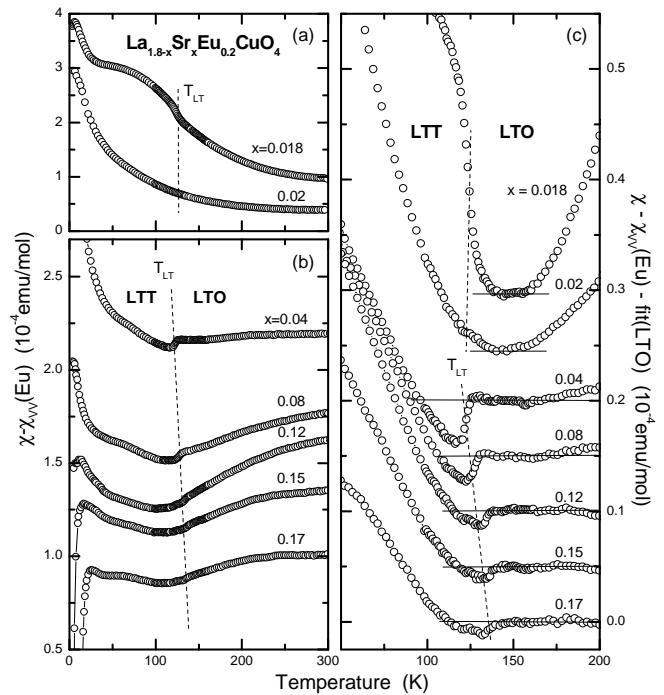}
\end{center} \vspace{-0.5cm} \caption[]{Static \sus\ ($H = 1$~Tesla) of
\lesco\ for different Sr concentrations $0.018\leq x \leq 0.17$ (a),(b)
after subtraction of $\chi_{VV}$(Eu) (data for $x=0.018$ and 0.02 the same
as in Fig.~\ref{c_smallx2_pap}). (c) after additional subtraction of
linear fit. In (a) curves are shifted by $5\times 10^{-5}$~emu/mol, in (b)
by $2.5\times 10^{-5}$~emu/mol, and in (c) by $5\times 10^{-6}$~emu/mol.}
\label{eu_msr_pap}
\end{figure}

\section{Overview of S\lowercase{r} and E\lowercase{u} doping dependence}
\label{doping}
In this section we present an overview of the evolution of $\chi -
\chi_{VV}$(Eu) as a function of Eu and Sr doping. All data in this section
are after subtraction of the Eu magnetism and were collected on
polycrystals. In Fig.~\ref{c_smallx2_pap}, we show data for \lesco\ with
variable Sr content $0 \leq x \leq 0.02$. Doping with Sr leads to a
continuous decrease of $T_N$, while the LT-transition remains at
$T_{LT}\sim 130$~K. In the LTO phase, the decrease of $T_N$ can easily be
followed by the position of the N\'eel peak. However, the N\'eel peak
disappears as soon as $T_N$ reaches the structural transition temperature,
which happens at a Sr content $x=0.016$. The size of the step in $\chi$ at
$T_{LT}$ is about the same as long as $T_N \gtrsim T_{LT}$, but then
decreases and eventually vanishes for $x=0.02$. This doping dependence
shows that the increase of $\chi - \chi_{VV}$(Eu) at the LT-transition is
connected with the presence of long range AF order.

Data for $x\geq 0.018$ are shown in Fig.~\ref{eu_msr_pap} (a) and (b).
Obviously, the step in \schi\ at $T_{LT}$ changes sign, then again
decreases and finally vanishes at high Sr concentrations $x\gtrsim 0.17$.
In Fig.~\ref{eu_msr_pap} (c) the data are shown on an enlarged scale after
subtraction of a linear fit applied immediately above the LT-transition.
In this figure one can clearly see the sign change of the anomaly at the
structural transition at about $x=0.02$.

\begin{figure}[t]
\begin{center}
\includegraphics[angle =0,width=0.39\textwidth,clip]{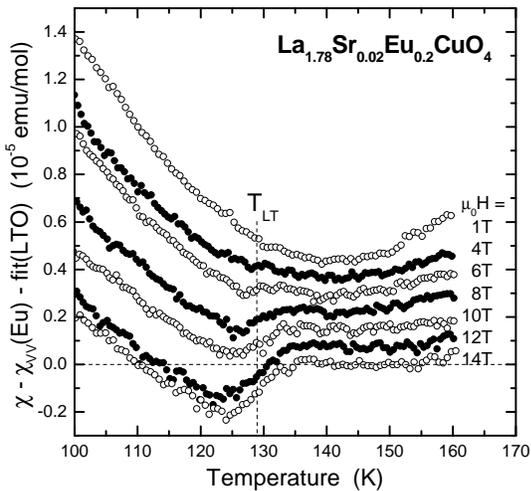}
\end{center} \vspace{-0.5cm} \caption[]{Static \sus\ of \leszco\ after
subtraction of $\chi_{VV}$(Eu) and linear fit for different magnetic
fields up to 14~Tesla. Curves are shifted for clarity by $0.75 \times
10^{-6}$~emu/mol.} \label{c_002h_pap}
\end{figure}

To obtain deeper insight, we have studied the special case $ x=0.02$ in
high magnetic fields. Corresponding data in Fig.~\ref{c_002h_pap} show
that with increasing field a negative anomaly of the same kind as for
$x>0.02$ is uncovered. We assume that at $x=0.02$ the positive jumps
observed for $x<0.02$ just compensate the negative jumps observed for
$x>0.02$. In Ref.~\onlinecite{Huecker97} we have argued that the negative
jumps for $x>0.02$ may signal an increase of the 2D magnetic correlation
length in the LTT phase. A similar jump in $\chi$ has been observed at the
charge stripe order transition in the nickelates.~\cite{Cheong94a} In the
nickelates, stripe order is not induced by a structural transition; i.e.,
the jump in $\chi$ is certainly connected to stripe order. In the cuprates
the situation is less clear since static stripe order occurs at much lower
temperatures than the LTO$\leftrightarrow$LTT transition.~\cite{Klauss00a}
Since the negative jump seems to vanish before the structural phase
boundary between the LTO and HTT phase at $x\simeq 0.25$ (for $T=T_{LT}$)
is reached~\cite{Klauss00a}, the size of the jump is possibly coupled to
the octahedral tilt angle.

\begin{figure}[t]
\center{\includegraphics[width=0.95\columnwidth,angle=0,clip]{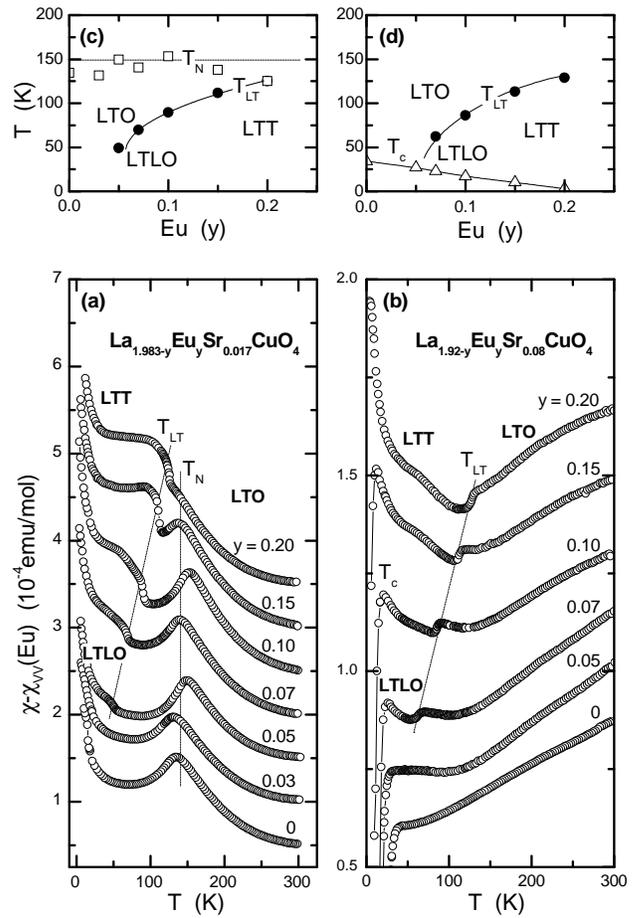}}
\caption[]{Static \sus\ of \lescoxy\ for fixed Sr concentration (a)
$x=0.017$ and (b) $x=0.08$ for different Eu doping $y$. (c) phase diagram
for $x=0.017$ and (d) for $x=0.08$.} \label{eu_Sr0017Sr008}
\end{figure}

In addition we have studied two series of samples as a function of the Eu
content $y$. The Sr content was fixed at $x=0.017$, which is in the AF
phase, and at $x=0.08$, which is in the underdoped superconducting phase.
For $x=0.017$ the LT-transition appears first at $y=0.05$ and
$T_{LT}\simeq 50$~K. With increasing Eu content, $T_{LT}$ as well as the
size of the jump in $\chi$ increases rapidly. At Eu content of $y=0.2$,
$T_{LT}$ reaches the magnetic transition at $T_N$ which appears to be
independent of the Eu content. As is well known, for small $y$ the
structural transition at $T_{LT}$ is of the LTO$\leftrightarrow$LTLO type
rather than of the LTO$\leftrightarrow$LTT
type.~\cite{Crawford91,Crawford93,Crawford97a,Buechner94c} We assume that
the increase of the jump in \schi\ with increasing $y$ is connected to a
gradual change of the transition from LTO$\leftrightarrow$LTLO to
LTO$\leftrightarrow$LTT. In the second series of samples with $x=0.08$,
the LT-transition appears first at $y=0.07$ and $T_{LT} \simeq 62$~K as a
small negative jump in \schi . With increasing $y$, $T_{LT}$ shifts to
higher temperatures. The size of the anomaly, however, does not change
significantly. Below $T_{LT}$ the slope $d\chi/dT$ shows a considerable
change with increasing $y$. Although for $T\gtrsim 70$~K this may
partially be due to a small inaccuracy in the subtraction of
$\chi_{VV}$(Eu), we emphasize that for $T\lesssim 70$~K the slope is real,
as $\chi_{VV}$(Eu) is constant. Fig.~\ref{eu_Sr0017Sr008}(b) also shows
that the onset temperature $T_c$ of the superconducting transition (at
1~Tesla) is suppressed from 34~K at $y=0$ to below 4~K at $y=0.2$. In
contrast, $\rm \mu$SR experiments have shown that, in the $\rm
Eu_{0.2}$-doped polycrystal, static magnetic (stripe) order occurs at
$T_N\simeq 6$~K, whereas in pure \lsco\ with x=0.08 this transition occurs
at the much lower temperature of 2.5~K.~\cite{Klauss00a,Niedermayer98} One
might wonder whether the steep increase of the \sus\ of the $\rm
Eu_{0.2}$-doped sample indicates the magnetic transition. However, no such
upturn is observed in our $\rm Eu_{0.2}$-doped single crystal with
$x=0.08$, showing that polycrystals in general seem to have a larger
number of magnetic defects.~\cite{Simovic03a} All critical temperatures
are summarized in Fig.~\ref{eu_Sr0017Sr008} (c) and (d). It is obvious
that in spite of the quite different Sr content, $T_{LT}(y)$ is almost
identical.

\section{Pure \lecoh}
\label{pure}
In this section we present the high-field magnetization of \leco\ ($x=0$)
obtained for a polycrystal (Sec.~\ref{poly}) and for a single crystal. The
polycrystal was annealed at 800$\rm ^\circ$C in vacuum ($T_N=316$~K). Its
\sus\ at 1~Tesla was already presented in Fig.~\ref{eu_ah88_pap_combi}.
The results for the single crystal annealed at 800$\rm ^\circ$C
($T_N=315$~K) will be presented in Sec.~\ref{SC1}, and those for the
crystal after annealing at 625$\rm ^\circ$C ($T_N=280~K$) in
Sec.~\ref{SC2}.

\begin{figure}[t]
\center{\includegraphics[width=1\columnwidth,angle=0,clip]{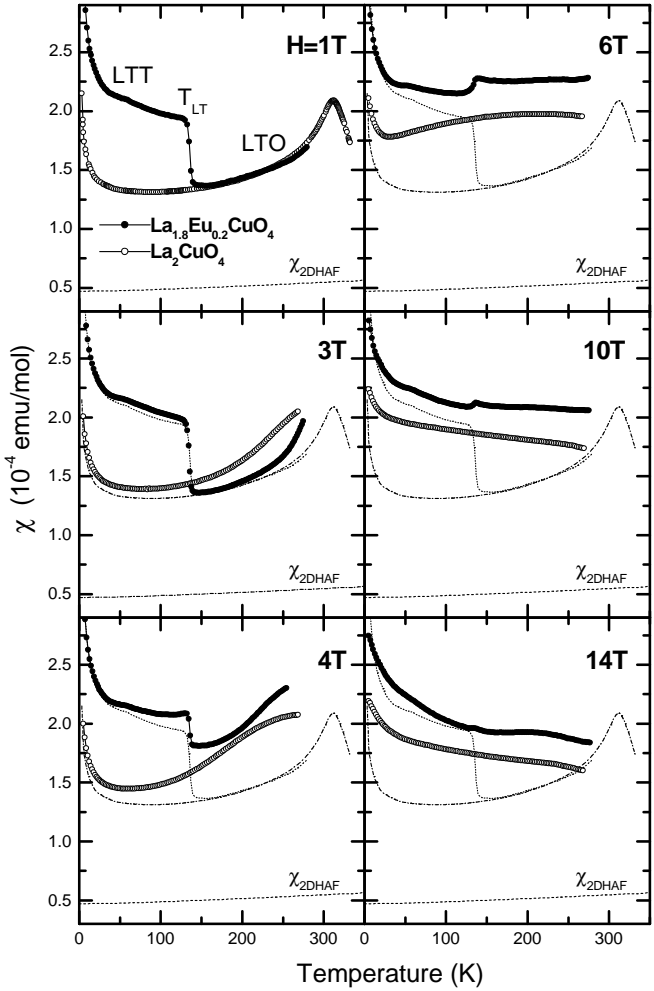}}
\caption[]{Static susceptibility of \leco\ after subtraction of
$\chi_{VV}$(Eu) and pure of \lco\ for different magnetic fields $H$. For
comparison 1~Tesla data shown in all plots. Data corrected for
$\chi_{dia}$ and $\chi_{VV}$(Cu) (cf. Fig.~\ref{eu_ah88_pap_combi}).
Dashed line: \sus\ of S1/2-2D-HAF for $J=1550$~K (after
Ref.~\onlinecite{Okabe88b})} \label{c_ah88_pap_neu}
\end{figure}

\subsection{The \lecohn\ polycrystal}
\label{poly}

\begin{figure}[t]
\center{\includegraphics[width=0.95\columnwidth,angle=0,clip]{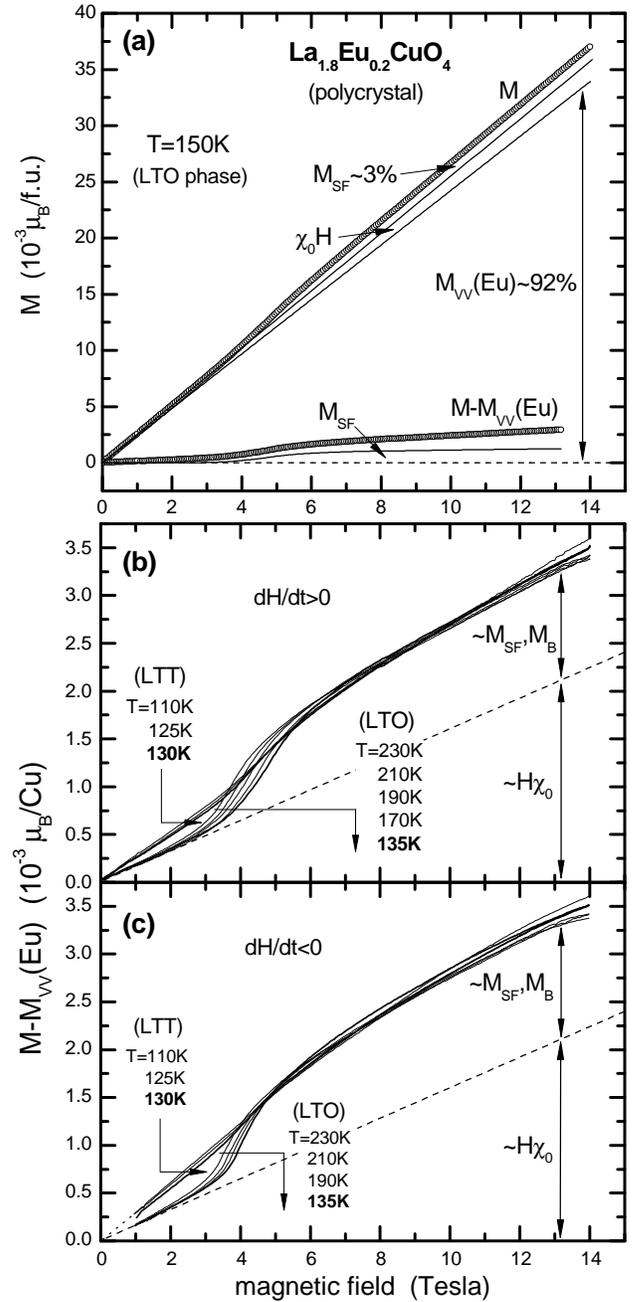}}
\caption[]{Magnetization $M(H)$ of \leco . (a) $M(H)$ and $M-M_{VV}$(Eu)
at 150~K. $M_{VV}$(Eu) is the calculated Van Vleck-term. (b),(c)
$M-M_{VV}$(Eu) at different temperatures from measurements with increasing
and decreasing magnetic field, respectively. $M_{SF}$ and $M_{B}$ indicate
spin-flip and weak ferromagnetic contribution of DM moment, respectively.}
\label{m_AH88_new_pap}
\end{figure}

In Fig.~\ref{c_ah88_pap_neu} we compare the \sus\ of the \leco\
polycrystal after subtraction of $\chi_{VV}$(Eu) and the \sus\ of a \lco\
polycrystal for magnetic fields up to 14 Tesla and temperatures up to
270~K. Data were corrected for $\chi_{dia}$ and $\chi_{VV}$(Cu) (cf.
Fig.~\ref{eu_ah88_pap_combi}). In the LTO phase, with increasing $H$, a
significant increase occurs for both samples. In contrast, in the LTT
phase of the Eu-doped compound changes are rather weak. In the LTO phase,
the influence of the field is particularly strong between 3 and 6~Tesla
and follows from the first order spin-flip transition that can be induced
for $H \parallel c$.~\cite{Thio88,Cheong89,Thio94,Huecker02b} As is shown
in Fig.~\ref{spinstructure1}~(a), at this transition the DM moments of
adjacent $\rm CuO_2$ planes, which are antiferromagnetically coupled in
zero magnetic field, become ferromagnetically
aligned.~\cite{Thio88,Kastner88} In the LTT phase, the spin-flip
transition is obviously very weak or absent. It is also interesting that
the jump at $T_{LT}$ drastically decreases, changes sign and almost
vanishes at 14~Tesla.~\cite{jump} Since in the LTO phase the critical
field $H_c$ increases with decreasing $T$, at intermediate fields of $\sim
4$~Tesla the spin-flip in \lco\ at high temperatures has progressed to a
higher degree than at low temperatures.~\cite{Cheong89} In contrast, at
maximum field the DM moments in \lco\ are ferromagnetically aligned down
to 4~K (to the degree that is possible in a
polycrystal~\cite{Huecker02b}). The \sus\ is smooth and increases
monotonically with decreasing $T$ and, moreover, is significantly larger
than $\chi_{2DHAF}$. A very similar temperature dependence is observed for
the Eu-doped sample at 14~Tesla. We therefore conclude that in \leco\ the
DM-moments are ferromagnetically aligned in both the LTO {\it and} the LTT
phases, though the higher \sus\ indicates that the DM-moments are even
larger than in \lco . Furthermore, the absence of a jump at $T_{LT}$ at
14~Tesla implies that the size of the DM-moment does not change
significantly at the structural transition LTO$\leftrightarrow$LTT. We
note that it is very difficult to understand this result in terms of a LTT
phase without DM spin canting as was suggested in
Ref.~\onlinecite{Tsukada03a}.

\begin{figure}[t]
\center{\includegraphics[width=1\columnwidth,angle=0,clip]{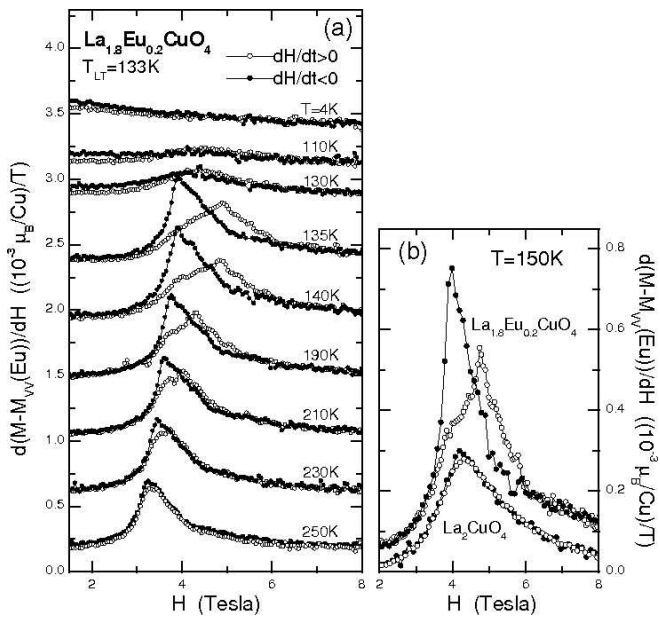}}
\caption[]{Derivative of $M-M_{VV}$(Eu) of \leco\ from  measurements with
increasing ($\circ$) and decreasing ($\bullet$) magnetic field. (a) in the
LTO and LTT phase. (b) at $T=150$~K in comparison with pure \lco . Curves
are shifted for clarity.} \label{m_AH88_deriv_pap}
\end{figure}

In Fig.~\ref{m_AH88_new_pap} we show a selection of $M(H)$ curves above
and below the LT-transition. Though in principle $\chi(T)$ provides
exactly the same information as $M(H)$, information about the spin-flip is
accessible more directly via $M(H)$. Extracting this information is,
however, a challenge. For example, in the case of the $M(H)$ curve at
150~K in Fig.~\ref{m_AH88_new_pap} (a), the magnetic contribution at
14~Tesla from the spin-flip amounts  to 3\% of the total signal, while
that of Eu accounts for 92\%.~\cite{precision} In
Fig.~\ref{m_AH88_new_pap}(b) and (c) we show $M-M_{VV}$(Eu) curves
measured with increasing and decreasing magnetic field at temperatures
above and below $T_{LT} \simeq 133$~K. The dashed line in this figure
represents the sum of all contributions that are approximately linear in
magnetic field. The contribution of the DM moments $M_{DM}$ in a
polycrystal is actually described by a function
$f(M_{DM})$.~\cite{Huecker02b} In the LTO phase we observe well defined
spin-flips; i.e., the magnetization shows a step like increase at the
critical spin-flip field $H_{c}$. The critical field increases with
decreasing temperature. In the LTT phase the spin-flip rapidly decreases
and in particular for $dH/dT<0$, it is almost absent for $T<130$~K.
However, this figure unambiguously shows that the magnetic contribution
due to the DM spin canting does not vanish in the LT phase. It rather
transforms from an antiferromagnetic-type to a ferromagnetic-type
magnetization of the DM moments.

The drastic changes that occur at $T_{LT}$, in particular between 135~K
and 130~K, are also clearly seen in the derivative $dM/dH$ shown in
Fig.~\ref{m_AH88_deriv_pap}(a). In the LTO phase, the spin-flip is
indicated by a pronounced maximum, which disappears in the LTT phase. This
figure also shows that $M(H)$ becomes strongly hysteretic below 250~K. In
contrast, in \lco\ the $M(H)$ curves stay reversible down to 150~K as is
displayed in Fig.~\ref{m_AH88_deriv_pap}(b) and was reported in
Ref.~\onlinecite{Huecker02b}. It is reasonable to attribute the enhanced
hysteretic behavior in the Eu-doped compound to a local lattice distortion
around the rare-earth site.

\subsection{\lecoh\ single crystal (800$\bf ^\circ$C vacuum)}
\label{SC1}

\begin{figure}[t]
\center{\includegraphics[width=0.95\columnwidth,angle=0,clip]{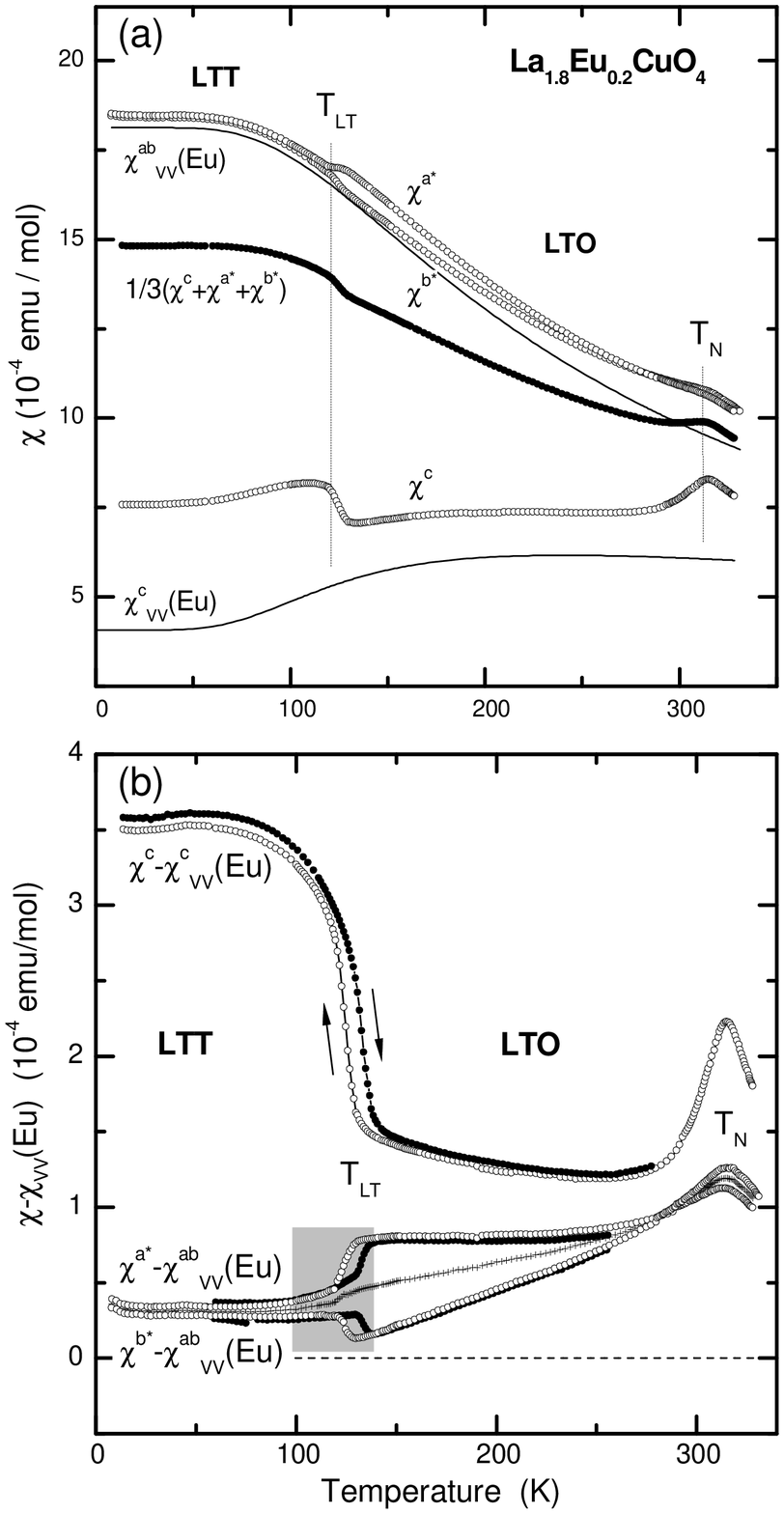}}
\caption[]{Static \sus\ ($H=1$~Tesla) of \leco\ for magnetic fields
perpendicular ($\chi^c$) and parallel to the $\rm CuO_2$ planes
($\chi^{a^*}$,$\chi^{b^*}$). (a) total signal from measurements with
decreasing temperature. (----) anisotropic Eu Van Vleck \sus . ($\bullet$)
polycrystal average. (b) after subtraction of $\chi_{VV}$(Eu) for
increasing ($\bullet$) and decreasing ($\circ$) temperature. (+) Average
of the two $ab$-measurements with decreasing $T$. In the shaded
temperature range a mixed phase of LTO, LTLO, and LTT may exist.}
\label{c_u6c}
\end{figure}

In the following we present the results on the \leco\ single crystal after
it was annealed at 800$^\circ$C in vacuum. In Fig.~\ref{c_u6c} we show the
total \sus\ [plot (a)] as well as the \sus\ after subtraction of the
anisotropic Eu Van Vleck magnetism [plot (b)]. The solid lines in
Fig.~\ref{c_u6c} (a) are the calculated Eu Van Vleck contributions. The Eu
fraction $y^*$ was chosen so that after subtraction of the Eu Van Vleck
term the resulting curves in Fig.~\ref{c_u6c} (b) are in fair agreement
with the \sus\ of pure \lco\ (cf. Ref.~\onlinecite{Sun91a,Lavrov01a}). In
this way we have determined the Eu content of the crystal to
$y=0.185(5)$.~\cite{eucontent} Note, that there is no Curie-like upturn of
\schi\ at low temperatures, documenting the absence of a significant
number of magnetic impurities, defect Cu spins or $\rm Eu^{2+}$ ions.

The two transitions at $T_N$ and $T_{LT}$ are clearly visible in all three
crystal directions (Fig.~\ref{c_u6c} (b)). The LT-transition shows a
temperature hysteresis of 10~K, i.e. $T_{LT}=134$~K for increasing $T$ and
124~K for decreasing $T$. The N\'eel peak and the jump at $T_{LT}$ are
largest for $H\parallel c$. For $H\parallel ab$ we find a similar in-plane
anisotropy as for \lesfco\ in Fig.~\ref{U2_J2zwei_pap3}(b), which largely
follows from the crystal field anisotropy of $\chi_{VV}$(Eu) in the LTO
phase. Largely means that (in contrast to $x=0.15$) for $x=0$ one has to
take into account that a small fraction of the $ab$-anisotropy comes from
the $\rm CuO_2$ planes.~\cite{Lavrov01a} The appearance of the
$ab$-anisotropy in the LTO phase clearly indicates that the crystal is
partially detwinned. Below $T_{LT}$ the $ab$-anisotropy abruptly decreases
to a finite value and then decreases continuously with decreasing
temperature. For $T\lesssim 100$~K the $ab$-\sus\ eventually becomes
isotropic within the error of the experiment, i.e. the structure becomes
LTT. (Note that the remaining anisotropy of $\sim 0.1\times
10^{-4}$emu/mol amounts to $\sim$0.5\% of the total signal, only.) The
finite $ab$-anisotropy in the intermediate temperature range $100~{\rm K}
< T < T_{LT}$ (shaded in) can have different reasons. One is that the
transition actually consists of a sequence of transitions:
LTO$\rightarrow$LTLO$\rightarrow$LTT. On the other hand one has to take
the temperature dependence of the LTO phase fraction into account.

At this point we mention that after the first annealing procedure at
625$\rm ^\circ$C the crystal was detwinned to a much higher degree, as one
can see from the apparently larger $ab$-anisotropy in Fig.~\ref{cu625_pap}
(a) (which we will discuss in detail later). Moreover, also the N\'eel
peak in Fig.~\ref{cu625_pap} shows a much stronger $ab$-anisotropy than in
Fig.~\ref{c_u6c} (b), indicating a lower degree of twinning, too. From
pure \lco\ it is well known that the N\'eel peak is largest for
$H\parallel c$ and $H\parallel b$ and that there is no peak for
$H\parallel a$.~\cite{Thio90,Sun91a,Lavrov01a} From these facts we infer
that, in the LTO phase, $\chi_{VV}^a$(Eu)$>\chi_{VV}^b$(Eu).
Note, that the average in-plane \sus\ ($+$) in Fig.~\ref{c_u6c} (b) does
not show a significant anomaly at $T_{LT}$. Obviously the jump occurs only
for $H\parallel c$, which indicates that it is connected to the DM spin
canting.

\begin{figure}[t]
\center{\includegraphics[width=1\columnwidth,angle=0,clip]{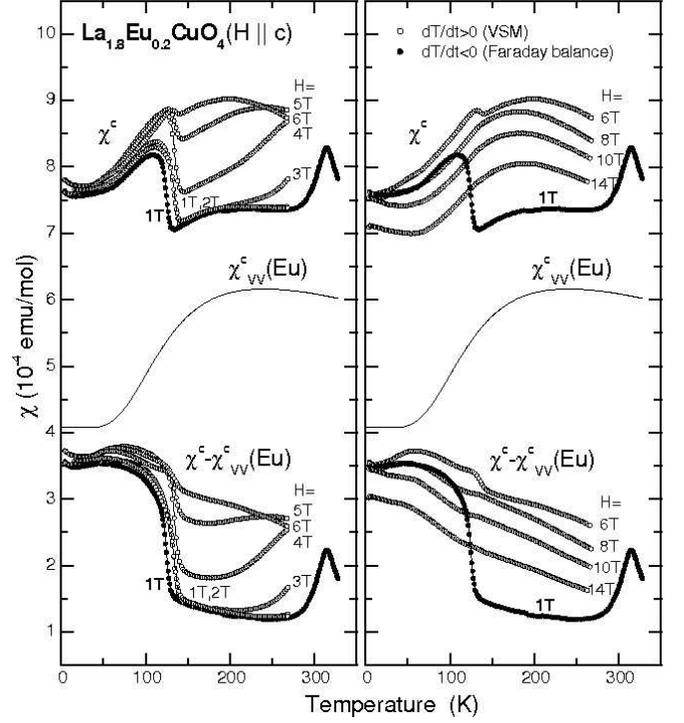}}
\caption[]{Static \sus\ $\chi^c$ of \leco\ before and after subtraction of
$\chi_{VV}^c$(Eu) for different $H$. Left: for $H\leq 6$~Tesla. Right: for
$H\geq 6$~Tesla. Measurements with VSM ($\circ$) were performed with
increasing $T$, for Faraday balance ($\bullet$) with decreasing $T$.}
\label{c_u6ch_pap}
\end{figure}

To study the DM spin canting we have performed measurements in high
magnetic fields with $H\parallel c$. Fig.~\ref{c_u6ch_pap} shows the total
\sus , the Eu Van Vleck term and the \sus\ after subtraction of this term.
Note that for $H\parallel c$ the ratio between the Cu spin magnetism and
the magnetic background due to the Eu ions is much more favorable than in
the polycrystal. Moreover, the jump at $T_{LT}$ is significantly larger,
since the signal is not averaged over all directions. It is as well about
twice as large as in Ref.~\onlinecite{Tsukada03a} for a single crystal
with $T_N=265$~K, underlining the strong impact of the excess-oxygen on
the intrinsic properties of the stoichiometric system (and hence the
importance of annealing under reducing conditions).

The left panel of Fig.~\ref{c_u6ch_pap} shows that there is a strong field
dependence in the LTO phase due to the spin-flip transition for $H\leq
6$~Telsa, which is similar to the polycrystal. In the LTT phase, the field
dependence is remarkably weak. The jump at $T_{LT}$ decreases, but does
not change sign, and finally vanishes around 8~Tesla. As is shown in the
right panel at higher fields, $H > 6$~Telsa, $\chi^c - \chi^c_{VV}$(Eu)
decreases in the LTO as well as the LTT phase. The smooth temperature
dependence of $\chi^c - \chi^c_{VV}$(Eu) at 14~Tesla, in particular around
$T_{LT}$, again shows that DM spin canting exists in the LTT phase and
that the size of the DM moment does not change at the transition.

\begin{figure}[t]
\center{\includegraphics[width=0.80\columnwidth,angle=0,clip]{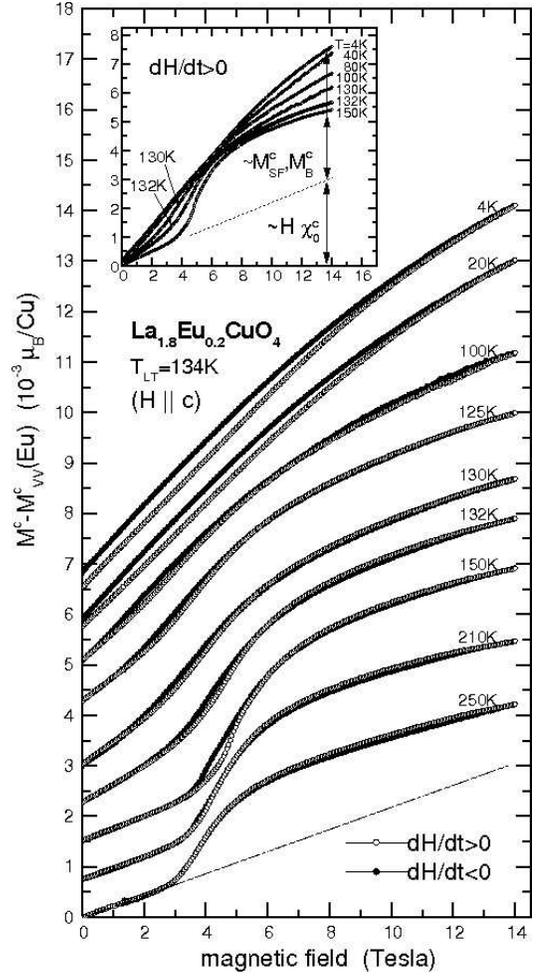}}
\caption[]{Magnetization $M^c-M^c_{VV}$(Eu) of \leco\ for increasing
($\circ$) and decreasing ($\bullet$) field. Data shifted for clarity.
Inset: unshifted data for increasing field. The dotted line approximately
indicates the contribution of $M^c_{SF}, M^c_{B}$ and $H\chi^c_0$.}
\label{m_euu6c_pap2}
\end{figure}

In Fig.~\ref{m_euu6c_pap2} we show a selection of $M^c(H)$ curves for
$H\parallel c$ for increasing as well as decreasing field. In the LTO
phase we observe well defined spin-flips, while below $T_{LT}$ they become
weaker and finally vanish. Similar to the polycrystal, the shape of the
$M^c(H)$ curves transforms from an antiferromagnetic-type to a weak
ferromagnetic-type. Moreover, at low temperatures $M^c(H)$ curves become
strongly hysteretic and show a significant remanent moment.

The inset shows non-shifted data for increasing field. The dashed line
approximates the linear magnetic background $H \chi_0^c$. Apparently, upon
decreasing the temperature the $M^c(H)$ curves in the LTT phase are the
envelope of the curves at higher temperature; i.e., the contribution of
the DM moments neither vanishes nor decreases. Note that in the LTO phase
the step in $M^c(H)$ due to the spin-flip is significantly larger than in
the case of the polycrystal, since in the crystal for $H\parallel c$ the
full DM moment can be aligned (see next paragraph). Another thing to
mention is that the slope $dM^c/dH$ at low magnetic fields significantly
increases for $T<T_{LT}$, which explains the jump in $\chi$(T) at
$T_{LT}$. Below $\sim 100$~K, however, $dM^c/dH$ becomes temperature
independent, i.e. it does not steepen further as is typical for a
classical ferromagnet. In particular at very low temperatures $M^c(H)$ is
almost linear up to $\sim 6$~Tesla before it bends in a ferromagnetic
fashion. In contrast, in \lndco\ at 2~K a ferromagnetic-type curvature is
already visible in $M(H)$ curves with a maximum field as low as 1.2~Tesla,
indicating that in this case it indeed seems to be the large 4$f$ moment
of Nd$^{3+}$ that shows a Brillouin-type magnetization.~\cite{Crawford93}

\begin{figure}[t]
\center{\includegraphics[width=1\columnwidth,angle=0,clip]{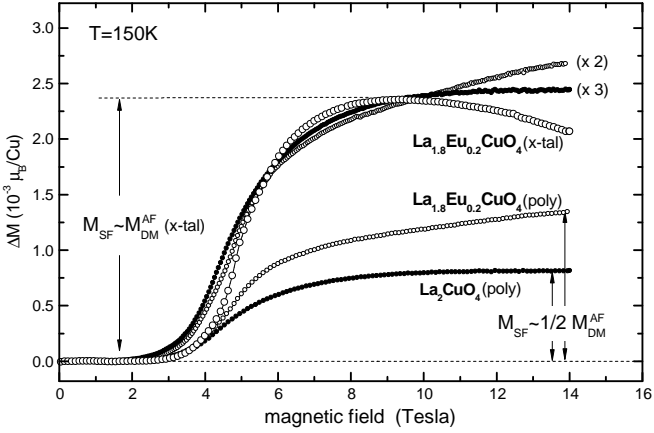}}
\caption[]{Deviations $\Delta M(H)$ of the magnetization from linearity at
0-2~Tesla in \leco\ single- and polycrystal as well as in \lco\
polycrystal at $T=150$~K. For comparison we show also scaled polycrystal
data.} \label{flipvgl}
\end{figure}

In Fig.~\ref{flipvgl} we show the deviations $\Delta M(H)$ from linearity
for the $M(H)$ curves at 150~K of the single crystal and of the Eu-doped,
as well as pure, \lco\ polycrystals. It is obvious that the single crystal
shows the largest step at the spin-flip transition. In an ideal single
crystal, the spin-flip causes a discontinous increase of $\Delta M$ by
$M_{DM}^{AF}$, where $M_{DM}^{AF}$ is the antiferromagnetically-ordered
fraction of the DM moment.~\cite{Thio88,Thio94} In a polycrystal, $\Delta
M$ increases continuously and converges to $\frac{1}{2} M_{DM}^{AF}$ in
the high field limit.~\cite{Huecker02b} If we multiply the data of the
Eu-doped polycrystal by a factor of 2 we indeed obtain a fair agreement
with the single crystal data. In the case of the \lco\ sample we have to
apply a factor of three. Consequently, in the Eu-doped sample the size of
the DM-moments is about 50\% larger. At 150~K a more precise analysis
yields $M_{DM}\simeq 2.45 \times 10^{-3}\mu_B$ for the Eu-doped single
crystal, $2.6 \times 10^{-3}\mu_B$ for the Eu-doped polycrystal and $1.8
\times 10^{-3}\mu_B$ for the \lco\ polycrystal (details in
Sec.~\ref{canting} and~\ref{diagrams}). It is well known that Eu-doping
causes the high temperature structural transition HTT$\leftrightarrow$LTO
to shift to higher temperature, which means that the octahedral tilt angle
at low temperature becomes larger.~\cite{Buechner94c} For a Eu content of
$y=0.2$ the HTT$\leftrightarrow$LTO transition shifts from $525$~K in pure
\lco\ to about $700$~K.~\cite{Werner00a}
To be more quantitative we have compared data from literature for
orthorhombic strain and octahedral tilt angles for pure and RE-doped \lco
.~\cite{Boeni88,Jorgensen88,Radaelli94,BradenDiss,JKlugeDipl,Crawford93,Crawford97a,Keimer93}
Thereafter, the apex oxygen tilt angle $\phi$ for $\rm Eu_{0.2}$-doping
compares to that for $\rm Nd_{0.35}$-doping, and, in the temperature range
$T_{LT} < T < 300$~K, is about $1.0(5)^\circ$ larger than in pure \lco ;
i.e., at 300~K $\phi$ increases from 4.2(5)$^\circ$ to 5.2(5)$^\circ$ and
at $T_{LT}{\rm (Eu_{0.2})}\simeq 135$~K from 5.1(5)$^\circ$ to
6.0(5)$^\circ$. It is reasonable to attribute the larger DM-moments to the
enlarged octahedral tilt angle. Note, however, that the relative change of
$\phi$ is small compared to that of the DM-moment, if we assume that
$M_{DM}\propto \phi$.~\cite{Thio88,Bonesteel93b} Though we have no
definite explanation for this discrepancy, we think that it might follow
from a non-linear relation between $M_{DM}$ and $\phi$ or from a doping
induced disorder of the octahedral tilts.

\begin{figure}[t]
\center{\includegraphics[width=1\columnwidth,angle=0,clip]{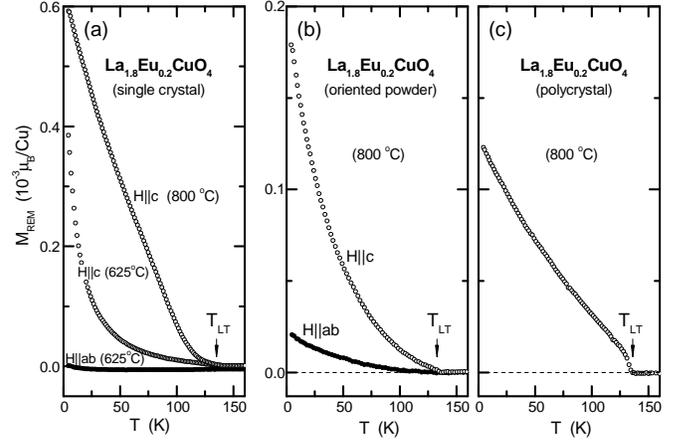}}
\caption[]{Remanent moment $M_{REM}=M(H=0)$ versus temperature of \leco\
after application of 14~Tesla at $T=4$~K for $H$ parallel and
perpendicular to $\rm CuO_2$ plane for (a) single crystal, (b) oriented
powder, and (c) polycrystal.} \label{u6_rem_pap}
\end{figure}

\begin{figure}[t]
\center{\includegraphics[width=0.7\columnwidth,angle=0,clip]{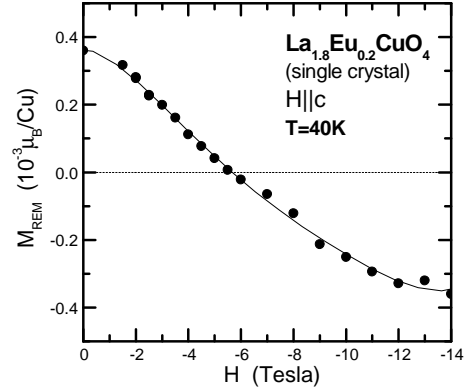}}
\caption[]{Remanent moment $M_{REM} \parallel c$ of \leco\ after
application of +14~Tesla at $T=4$~K as a function of a negative
counter-field at $T=40$~K.} \label{invrem_pap}
\end{figure}

In Ref.~\onlinecite{Shamoto92,Crawford93} it was reported that in \lndco\
and \lsmco\ magnetization curves exhibit a remanent moment below $T_{LT}$.
However, it was not clear whether these remanent moments stem from the
polarized $4f$-moments of the Nd and Sm ions due to an interaction with
the Cu spins.~\cite{Stein96}
Corresponding results for our Eu-doped samples are summarized in
Fig.~\ref{u6_rem_pap} where we show the remanent moment $M_{REM}$ at $H=0$
after application of 14~Telsa at 4~K as a function of increasing
temperature for the single crystal (left), the polycrystal (right) and an
oriented powder (middle) made from the polycrystal. In all three cases we
observe a remanent moment which decreases with increasing $T$ and
disappears at $T_{LT}$. The results on single crystal and oriented powder
clearly show that $M_{REM}$ is perpendicular to the $\rm CuO_2$ planes,
which supports the idea that $M_{REM}$ is caused by ferromagnetically
aligned DM-moments. Note, that the alignment of the powder was successful
to about ~80\% which is the reason for the small remanent moment for
$H\parallel ab$. At 4~K $M_{REM}$ in single- and polycrystal amount to
about 10-20\% of $M_{DM}$, only.

In Fig.~\ref{invrem_pap} we display $M_{REM}$ of the single crystal at
40~K as a function of a negative counter-field after $+14$~Tesla was
applied at 4~K. Each time after the negative counter-field was increased,
the remaining $M_{REM}$ was measured at zero field. The initial idea of
this measurement was to check if there is a particular counter-field at
which the remanent moment can be inverted. Obviously this is not the case.
$M_{REM}$ is inverted monotonically and $-14$~Tesla has to be applied to
fully invert the remanent moment initially induced by $+14$~Tesla.

\subsection{\lecoh\ single crystal ($\bf 625$ $\bf ^\circ$C $\bf N_2$)}
\label{SC2}

\begin{figure}[t]
\center{\includegraphics[width=1\columnwidth,angle=0,clip]{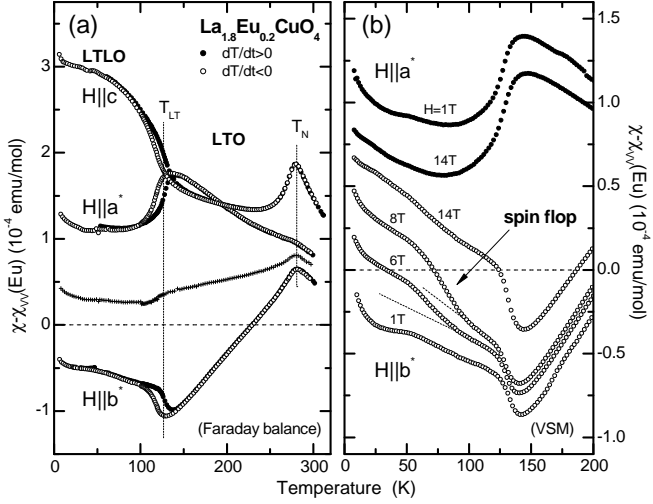}}
\caption[]{Static \sus\ of \leco\ after subtraction of $\chi_{VV}$(Eu).
(a) for $H=1$~Tesla for all three crystal directions and increasing
($\bullet$) as well as decreasing ($\circ$) temperature. (+) Average of
the two $ab$-measurements with decreasing $T$. (b) for $H$ up to 14~Tesla
parallel to the two $ab$-directions as a function of increasing $T$.}
\label{cu625_pap}
\end{figure}

In Fig.~\ref{cu625_pap}(a) we display $\chi - \chi_{VV}$(Eu) at 1~Tesla
for the \leco\ single crystal after its first annealing at $ 625^\circ$ in
$N_2$. Note that for $H\parallel ab$ the average $ab$-plane Eu$^{3+}$ Van
Vleck \sus\ has been subtracted. Aside from the lower N\'eel temperature
of 280~K, one recognizes the broader as well as smaller jump at $T_{LT}$
for $H\parallel c$. Moreover, the $ab$-anisotropy in the LTO phase does
not disappear at low $T$, which is a clear sign that the low-temperature
structure is of the LTLO-type and not of the LTT-type. In fact, the
reduction of the $ab$-anisotropy by $\sim$40\% from $\sim 2.8\times
10^{-4}$~emu/mol immediately above $T_{LT}$ to $\sim 1.6\times
10^{-4}$~emu/mol at 50~K indicates that the azimuthal rotation of the
octahedral tilt axis is about 27$^\circ$ only, which is significantly
smaller than $45^\circ$. As mentioned previously the absence of a
pronounced N\'eel peak for $H\parallel a^*$ indicates that the crystal is
almost perfectly detwinned, where $a^*$ ($b^*$) is the crystal direction
predominantly containing the $a$-axis ($b$-axis). This provides the
opportunity to study the in-plane anisotropy in more detail. In
Fig.~\ref{cu625_pap}(b) we plot $\chi-\chi_{VV}$(Eu) for different
magnetic fields. Whereas for $H\parallel a^*$ the \sus\ decreases slightly
between 1 and 14~Tesla, we observe a significant and uneven increase for
$H\parallel b^*$. In particular in the LTLO phase we observe a magnetic
transition which we attribute to a spin-flop. In \lco\ the spin-flop
occurs for $H\parallel b$ and is a measure of the in-plane spin-wave
gap.~\cite{Thio90}

\begin{figure}[t]
\center{\includegraphics[width=0.98\columnwidth,angle=0,clip]{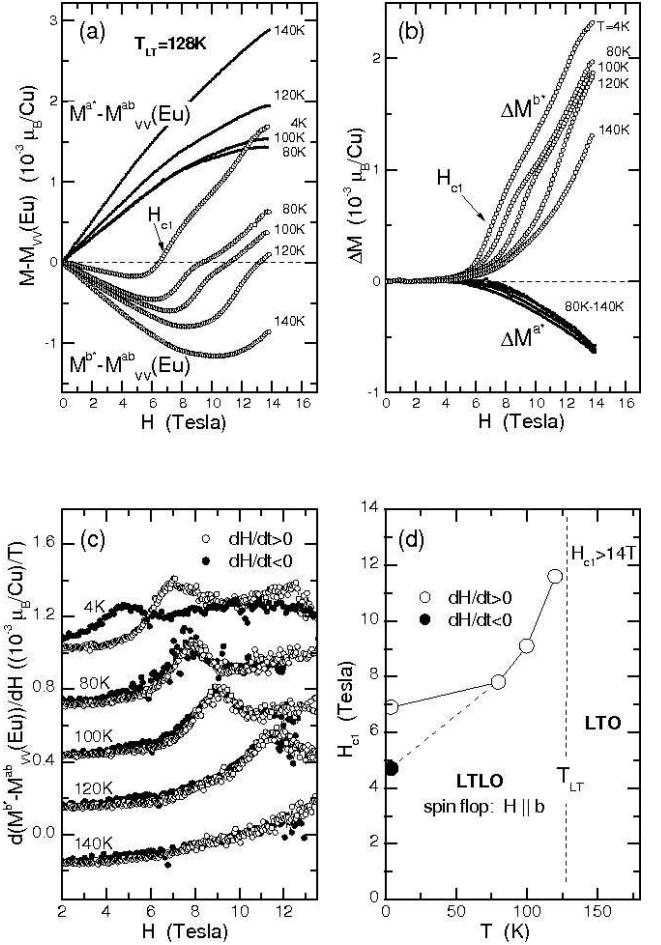}}
\caption[]{Spin-flop in \leco . (a) magnetization $M^{a^*}$ and $M^{b^*}$
of \leco\ after subtraction of Eu magnetism for different temperatures in
the LTO and LTLO phase. (b) deviation of $M^{a^*}$ and $M^{b^*}$ from
linearity. (c) derivative $d(M^{b^*}-M_{VV}^{ab}{\rm (Eu)})/dH$ at
different temperatures. Curves are shifted for clarity. (d) spin-flop
field $H_{c1}$ versus temperature.} \label{mu625neu_pap}
\end{figure}

Corresponding $M(H)$ curves in Fig.~\ref{mu625neu_pap} yield a more
detailed picture of this transition. It is sufficient to focus on plot (b)
where we show the deviations from linearity of the $M(H)$ curves in plot
(a). In accordance with the \sus\ data the magnetization shows a weak
negative curvature for $H\parallel a^*$ and a positive curvature with
clear magnetic transitions for $H\parallel b^*$. Our data for $H\parallel
b^*$ behave qualitatively similarly to a measured, as well as calculated,
spin-flop for \lco\ in Fig.~3 and Fig.~9(a) of Ref.~\onlinecite{Thio90},
respectively. We mention that the measured transition is much broader than
the calculated one. One reason for this is certainly the excess-oxygen. In
Ref.~\onlinecite{Thio90} it is argued that the broadening can occur
because the magnetic field is not exactly parallel to the
$b$-axis.~\cite{spinflop} Thus in the LTLO phase we have to expect a
broadening of the spin-flop, since the spins follow the azimuthal rotation
of the octahedral tilt axis; i.e., spins intrinsically rotate out of the
direction of the $b$-axis [cf. Fig.~\ref{spinstructure1}~(b)]. However, it
is important to notice that if field and spins are off-axis, this should
mainly lead to a broadening and not to a change of the critical spin-flop
field $H_{c1}$.~\cite{Thio90} From this perspective it is very interesting
that below $T_{LT}$ the critical field $H_{c1}$ strongly decreases and
becomes hysteretic at low $T$ as can be seen from the derivative
$d(M^{b^*}-M^{ab}_{VV}{\rm(Eu)})/dH$ and the determined phase diagram in
Fig.~\ref{mu625neu_pap}.~\cite{spinflop} Since $H_{c1}$ is a measure of
the in-plane gap, our results suggest that in the LTLO phase the effective
in-plane gap decreases, which is in contradiction to the increase reported
in Ref.~\onlinecite{Keimer93} (details in discussion).

\subsection{Analysis of spin canting in LTO and LTT phase}
\label{canting}
For a more quantitative analysis of $M_{DM}$ we have analyzed the $M^c(H)$
curves of the single crystal (after it was annealed at 800~$^\circ$C). In
Fig.~\ref{LTTdyn} we show the fit results for two $M^c(H)$ curves measured
above (left) and below (right) the LT-transition according to the
function:
\begin{figure}[t]
\begin{center}
\includegraphics[angle =0,width=0.47\textwidth,clip]{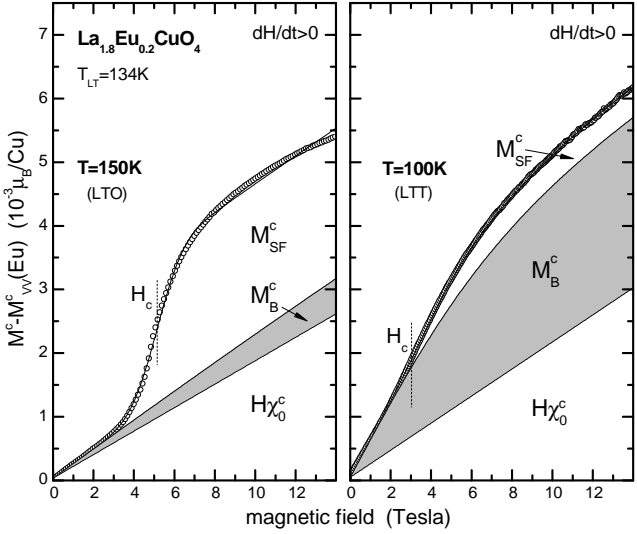}
\end{center}
\vspace{-0.5cm} \caption[]{Magnetization $M^c(H)$ of \leco\ after
subtraction of Eu Van Vleck term $M^c_{VV}$(Eu). Left: in the LTO phase.
Right: in the LTT phase. (-----) result of fit according
Eq.~\ref{equ_hsfsc} as well as the various contributions $H\chi_0^c$,
$M^c_{SF}$ und $M^c_{B}$.} \label{LTTdyn}
\end{figure}
\begin{equation}\label{equ_hsfsc}
M^c - M^c_{VV}(Eu) = H\cdot \chi^c_0 + M^c_{SF} + M^c_{B}\vspace{0.3cm}
\end{equation}
where $H \chi^c_0$ represents all terms that are linear in field,
$M^c_{SF}$ describes the spin-flip of the antiferromagnetically ordered
part of the DM-moments $M_{DM}^{AF}$, and $M^c_{B}$ the contribution of
that part of the DM-moment $M_{DM}^{WF}$ which behaves like a weak
ferromagnet. $M^c_{SF}$ is described by a step of the size of
$M_{DM}^{AF}$ with a gaussian distribution $\Delta H_c$ of the critical
field $H_c$:
\begin{equation} M^c_{SF} = M_{DM}^{AF}\frac{1}{\sqrt{2\pi}\Delta H_c} \int_{0}^{H} e^{-\frac{(H^*-H_c)^2}{2\Delta H_c^2}} dH^*
\end{equation}
and $M^c_{B}$ by a Brillouin function for spin 1/2:
\begin{equation}
M^c_{B}=M_{DM}^{WF} {\rm tanh}(k H/T)
\end{equation}
where $k$ is a fit parameter. The $M^c(H)$ curves for sixteen temperatures
were fitted simultaneously. The linear term $H \chi^c_0$ was confined to
vary linearly as a function of $T$. All other fit parameters
$M_{DM}^{AF}$, $M_{DM}^{WF}$, $H_c$, $\Delta H_c$, and $k$ were varied
independently for each $M^c(H)$ curve. Overall the fits provide a good
description of the data (solid lines in Fig.~\ref{LTTdyn}).
\begin{figure}[t]
\begin{center}
\includegraphics[angle =0,width=0.48\textwidth,clip]{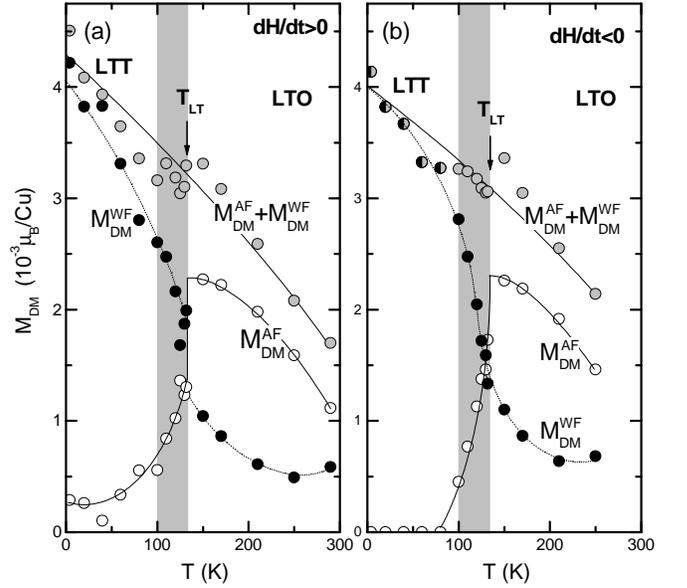}
\end{center} \vspace{-0.5cm} \caption[]{Antiferromagnetic $M_{DM}^{AF}$ and weak
ferromagnetic $M_{DM}^{WF}$ parts of DM moment as well as their sum
$M_{DM}^{AF} + M_{DM}^{WF}$ in the \leco\ single crystal as a function of
temperature for (a) increasing and (b) decreasing field. In the shaded
temperature range a mixed phase of LTO, LTLO, and LTT may exist in \leco .
Solid lines are guides to the eye.} \label{U6_MDM_pap}
\end{figure}
The comparison in Fig.~\ref{LTTdyn} shows that at 150~K (LTO) the
spin-flip contribution $M^c_{SF}$ is significantly larger then the weak
ferromagnetic contribution $M^c_{B}$.~\cite{para} In contrast, at 100~K
(LTT) it is $M^c_{B}$ which dominates over $M^c_{SF}$ and the critical
field $H_c$ of the residual spin-flip has shifted to lower values.

In Fig.~\ref{U6_MDM_pap} we plot the temperature dependence of
$M_{DM}^{AF}$, $M_{DM}^{WF}$ and $M_{DM}^{AF} + M_{DM}^{WF}$ we have
extracted from $M^c(H)$ curves measured with increasing and decreasing
field. At the transition into the low-temperature phase $M_{DM}^{AF}$
drastically decreases, while $M_{DM}^{WF}$ increases. However, the sum of
both terms, within the error of our analysis, increases monotonically. In
particular, there is no drastic change of the DM moment at the structural
transition itself:
\begin{equation}\label{equ_LTTdyn}
\begin{array}{ll}\vspace{-0.1cm}
[M_{DM}^{AF}+M_{DM}^{WF}](LTT)_{_{T\lesssim T_{LT}}} & \vspace{0.5cm}\\
 & \hspace{-3cm} \simeq [M_{DM}^{AF}+M_{DM}^{WF}](LTO)_{_{T\gtrsim T_{LT}}}\ .
\end{array}
\end{equation}
At $T=4$~K we find a DM moment of $\simeq 4 \times 10^{-3} \rm \mu_B$/Cu
which is about 50\% larger than in pure \lco , in agreement with our
comparison of $M(H)$ at 150~K in Fig.~\ref{flipvgl}. Obviously DM spin
canting does not disappear in the LTT phase. There is, however, a
significant shift of weight from $M_{DM}^{AF}$ to $M_{DM}^{WF}$.

\begin{figure}[t]
\begin{center}
\includegraphics[angle =0,width=0.31\textwidth,clip]{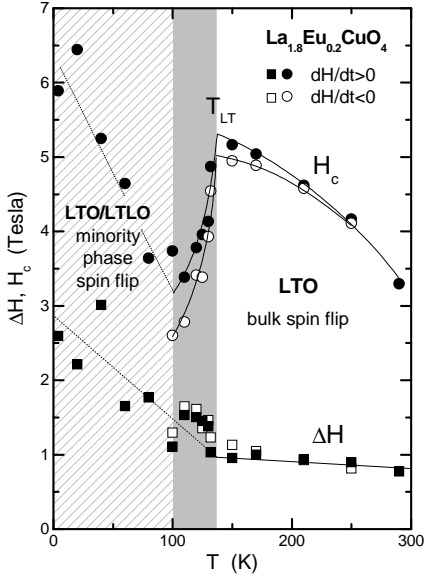}
\end{center} \vspace{-0.5cm} \caption[]{Spin-flip field $H_c$ and width
of the transition $\Delta H_c$
of the \leco\ single crystal as a function of temperature for increasing
(full symbols) and decreasing (open symbols) field. Solid lines are guides
to the eye.} \label{U6_HCDH_pap}
\end{figure}

The shaded regions in Fig.~\ref{U6_MDM_pap} mark the temperature range
$100~{\rm K} < T < T_{LT}$ where, based on the \sus\ in Fig.~\ref{c_u6c}
(b), the crystal, after the discontinuous decrease at $T_{LT}$, still
shows a clear $ab$-anisotropy. It is reasonable to assume that in this
region volume fractions of LTO, LTLO, and LTT coexist (cf.
Sec.~\ref{experimental}). In this temperature range the remaining
spin-flip is clearly visible for increasing as well as decreasing field.
In contrast, no well defined spin-flip is observed for $T< 100$~K. Here,
weak signatures ($M_{DM}^{AF}\ll 1\times 10^{-3}{\rm \mu_B}$/Cu) are
detected only for increasing field while there is no sign
($M_{DM}^{AF}=0$) of a spin-flip for decreasing field (see also
Fig.~\ref{m_euu6c_pap2}). We do not believe that the spurious spin-flip
for increasing field represents a bulk property of the LTT phase. We
rather think that it is connected to a LTLO or LTO minority phase. This
interpretation is supported by the temperature dependence of the critical
field $H_c$ in Fig.~\ref{U6_HCDH_pap}. While $H_c$ drops sharply for
$100~{\rm K}<T<T_{LT}$, it again starts to increase for $T<100$~K. First
of all, this would be a very unusual temperature dependence if it were
intrinsic to LTT. Moreover, at low temperatures $H_c$ reaches $\sim
6$~Tesla, which is roughly the zero-temperature approximation for the LTO
phase. At the same time, the width $\Delta H$ increases from 1~Tesla in
the LTO phase to about $\sim 2.5$~Tesla, underlining that this transition
is not at all well defined. Hence, our main conclusion with regard to the
bulk properties of \leco\ is that spin-flips with well defined $H_c$ occur
only in the LTO phase, whereas no spin-flip exists in the LTT phase.

\subsection{Phase diagrams of \lecoh}
\label{diagrams}
\begin{figure}[t]
\center{\includegraphics[width=0.81\columnwidth,angle=0,clip]{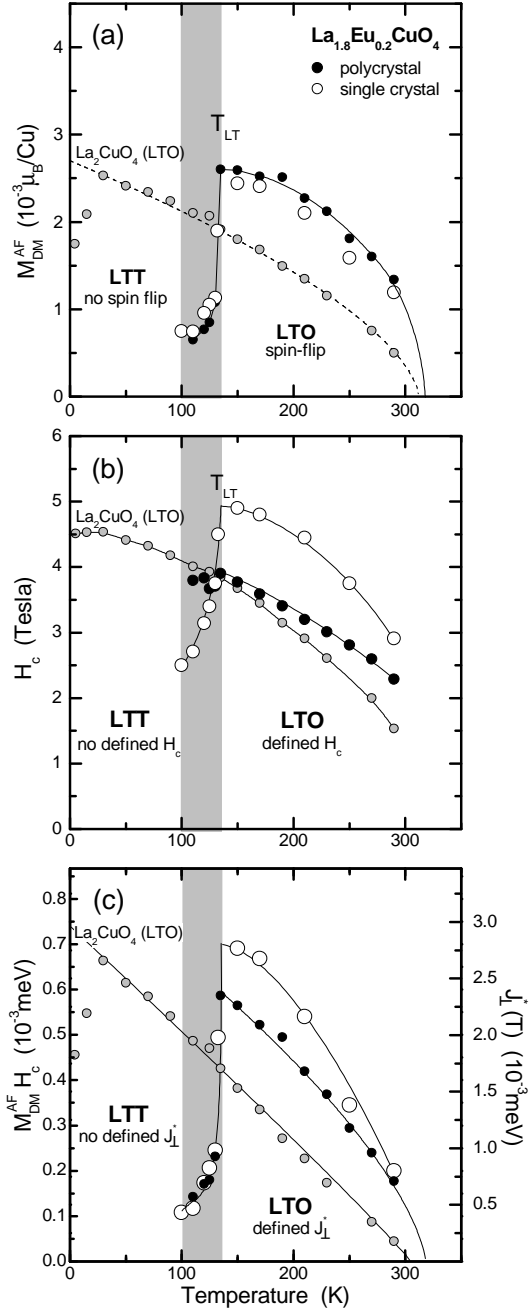}}
\caption[]{Spin-flip parameters: (a) antiferromagnetically ordered part
$M_{DM}^{AF}$ of DM moment, (b) spin-flip field $H_c$, and (c) effective
interlayer coupling $J_\perp^{*}\propto M_{DM}^{AF}\times H_c$ versus
temperature in \leco\ single- and polycrystal and in the \lco\
polycrystal. In the shaded temperature range a mixed phase of LTO, LTLO,
and LTT may exist in \leco .} \label{euhcneu_pap}
\end{figure}

In Fig.~\ref{euhcneu_pap} we compare the temperature dependence of
$M_{DM}^{AF}$, $H_c$, and $M_{DM}^{AF} \times H_c$ for the \leco\ single
and polycrystal as well as the \lco\ polycrystal. The $M(H)$ curves of the
polycrystals had to be fitted separately for every temperature, since, due
to the more complex fit function, simultaneous fits were not
stable.~\cite{Huecker02b} Also, $H \chi_0$ was varied independently. For
the sake of consistency, we have refitted the single crystal data
accordingly. For $M_{DM}^{AF}$ in Fig.~\ref{euhcneu_pap} (a) we find a
convincing agreement between the \leco\ single and polycrystal.
Moreover, the comparison with \lco\ shows the much larger values for
$M_{DM}^{AF}$ in the LTO phase of the Eu-doped compounds. Again, we
emphasize that $M_{DM}^{AF}$ is the AF ordered part of the DM moment only,
and not a measure for the full DM moment.

$H_{c}$ in Fig.~\ref{euhcneu_pap} (b) was determined from the average for
measurements with increasing and decreasing fields. In the LTO phase,
$H_c$ increases with decreasing $T$ and is significantly higher in the
single crystal than in either of the polycrystals. A possible explanation
may involve finite size effects in the polycrystals due to a limited
magnetic correlation length in the crystallites.
On the other hand, we find for the single crystal that, with increasing
oxygen content, the spin-flip transition broadens, and the maximum in
$dM^c/dH$ associated with $H_c$ effectively shifts to \textit{higher}
fields. Therefore, the crystal's relatively high $H_c$ values in the LTO
phase might to some extent follow from a finite and possibly inhomogeneous
oxygen excess. Further experiments are needed to clarify this problem.

Below $T_{LT}$, the spin-flip field decreases, for the single crystal in
particular. In the polycrystal, $M(H)$ changes so drastically below
$T_{LT}$ that the average $H_c$ can only be followed down to 110~K. In
this temperature range of $\sim$25~K below $T_{LT}$, $H_c$ drops only
slightly. Since the structural transition in the polycrystal is much
sharper than in the single crystal, we infer that the polycrystal
approaches the LTT phase at a higher temperature than the single crystal
(cf. Sec.~\ref{experimental}); i.e., the polycrystal is closer to showing
the intrinsic properties of the LTT phase. The residual spin-flip,
therefore, can be associated with a minority phase LTO and/or LTLO whose
volume fraction rapidly decreases with decreasing $T$. The fact that $H_c$
below $T_{LT}$ remains almost as high as in the LTO phase supports this
interpretation.

In Fig.~\ref{euhcneu_pap} (c) we show the temperature dependence of the
effective interlayer coupling:
\begin{equation}\label{inter}
J_\perp^*(T)=M_{DM}^{AF}(T)\cdot H_c(T)/S^2
\end{equation}
which has been introduced and discussed in detail in
Ref.~\onlinecite{Huecker02b}. $J_\perp^*$ is a measure for the strength of
the AF interlayer coupling at a particular temperature and only in the
limit $T \rightarrow 0$ identical with the interlayer superexchange
constant $J_\perp$.~\cite{Thio88} It is apparent from Eq.~\ref{inter} that
$J_\perp^*(T)$ is an implication of $M_{DM}^{AF}(T)$ and $H_c(T)$. In the
LTO phase of the Eu-doped samples $J_\perp^*$ is well-defined and is
significantly higher than in pure \lco. Below the transition, $J_\perp^*$
decreases by $\sim$80\% between $T_{LT}$ and 100~K. At temperatures
$T<100$~K no well-defined $J_\perp^*$ exists. In fact we assume that in
the LTT phase $J_\perp^*$ is not defined at all. Since we associate the
remains of the spin-flip below $T_{LT}$ with a LTLO and/or LTO phase with
rapidly decreasing volume fractions, the changes of $J_\perp^*$ in the
shaded region might be much smaller than suggested by
Fig.~\ref{euhcneu_pap} (c), since $M_{DM}^{AF}$ has to be normalized with
the volume fraction.

Let us go back to the strong reduction of $M_{DM}^{AF}$ below $T_{LT}$.
This observation shows us that the $M(H)$ curves in the LTT phase do not
comply with a simple shift of a spin-flip of unchanged magnitude to a
lower critical field. This means that there is no simple, uniform
reduction of the interlayer coupling $J_\perp^*$. On the other hand we do
not observe spontaneous weak ferromagnetism which indicates that
$J_{\perp}^*$ is not zero, either. In the case $J_\perp^* = 0$ we would
expect that the full DM moment can be ferromagnetically aligned by a
relatively small field which would lead to a large initial slope $dM/dH$.
We mention that Vierti\"o and Bonesteel~\cite{Viertio94a} have calculated
$M(H)$ curves for different interlayer coupling mechanisms in the LTO and
LTT phases and that weak ferromagnetism in the LTT phase is one of their
solutions. Obviously, our measured $M(H)$ curves do not show this type of
behavior (as well as none of the other types).
Nevertheless, the fact that in the LTT phase $dM/dH$ at low fields is
significantly larger than in the LTO phase indicates that $J_\perp^*$ is
reduced (cf. Fig.~\ref{m_euu6c_pap2} and \ref{m_AH88_new_pap}). Thus, we
believe that the LTT phase in our sample is characterized by a broad
distribution of $J_\perp^*$ with the center of mass shifted to lower
values than in the LTO phase. This idea is supported by the measurement of
$M_{REM}$ in Fig.~\ref{invrem_pap} which shows that, on the one hand,
there is no well defined critical field to invert $M_{REM}$, and on the
other hand, that the required field to fully invert $M_{REM}$ is rather
large. Nonetheless, it still remains unclear to us whether a broad
distribution of $J_\perp^*$ represents the intrinsic properties of an
ideal strain-free LTT phase.~\cite{strain}\\

\section{Lightly S\lowercase{r} doped \lecoh}
\label{lightly}
\subsection{Comparison with \lscoh}
\begin{figure}[t]
\center{\includegraphics[width=1\columnwidth,angle=0,clip]{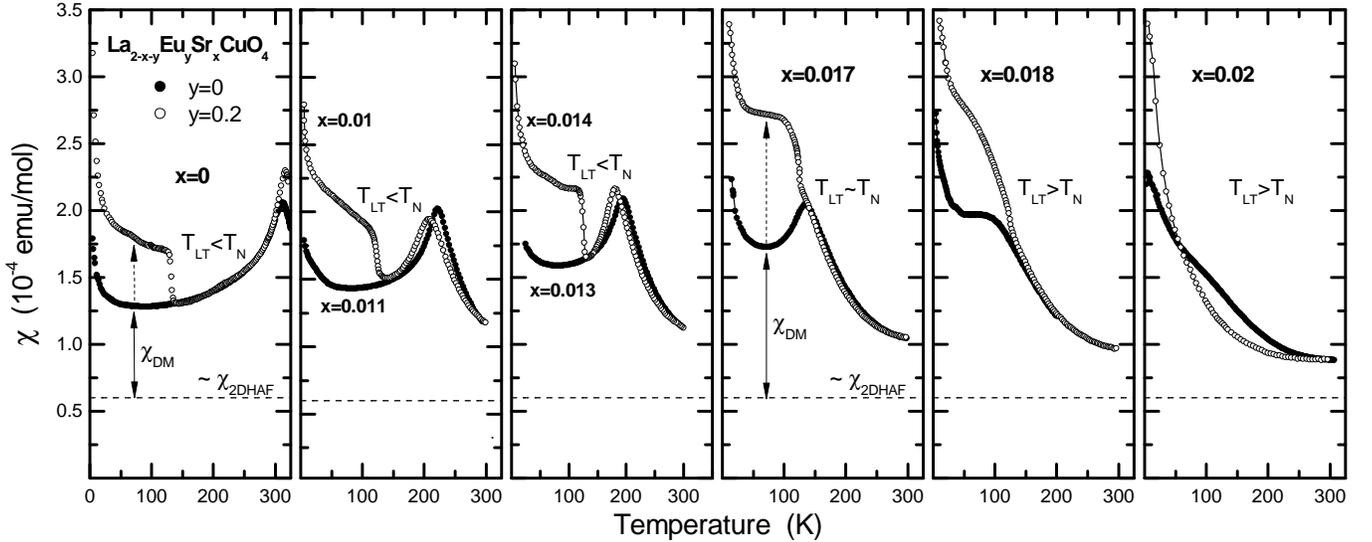}}
\caption[]{Comparison of $\chi$ at 1~Tesla of \lesco\ after subtraction of
$\chi_{VV}$(Eu) and pure \lsco\ for different Sr concentrations $x<0.02$.
Data also corrected for $\chi_{dia}$ and $\chi_{VV}$(Cu) (cf.
Fig.~\ref{eu_ah88_pap_combi}). The arrows for $x=0$ and 0.017 indicate the
contribution of the DM-Moments $\chi_{DM}$ in the LTO and LTT phase. The
dashed line approximately represents $\chi_{2DHAF}$.} \label{eu02eu00_pap}
\end{figure}

In this section we study the influence of light Sr doping ($x\leq 0.02$)
on the Cu spin magnetism in the LTT phase. For this purpose, we directly
compare in Fig.~\ref{eu02eu00_pap} the \sus\ $\chi$ of pure and $\chi -
\chi_{VV}$(Eu) of Eu-doped \lsco\ for polycrystalline samples with similar
Sr content and $T_N$.~\cite{Kataev99b} The Eu fraction $y^*$ was varied,
so that $\chi - \chi_{VV}$(Eu) and $\chi$ match at high temperatures (for
$x=0$ at 150~K). The determined $y^*$ values agreed with the nominal value
$y=0.2$ within the experimental error of $\Delta y\simeq 0.005$.
Furthermore, all data sets were corrected for $\chi_{dia}+\chi_{VV}$(Cu),
so that they can be compared to $\chi_{2DHAF}$, which we have approximated
by the dotted lines. As mentioned before, the difference between the data
and $\chi_{2DHAF}$ is caused by the DM spin canting.

In the LTO phase, the pairs of curves show a fair agreement in the
paramagnetic phase as well as in the AF phase. In contrast, the
differences between the LTT and LTO phase at Sr content $x=0$ become even
larger with increasing $x$. This result shows that DM spin canting is
present in the LTT phase of Sr-doped samples, as well. Eventually, for $x
\gtrsim 0.017$ ($T_N \lesssim T_{LT}$) the differences between LTT and LTO
start to decrease.

As was shown in Ref.~\onlinecite{Huecker02b}, the effective interlayer
coupling in pure \lsco\ strongly decreases with increasing Sr content and
for $x=0.017$ is already quite weak. In the Eu-doped compounds, the
LT-transition causes a further reduction of the interlayer coupling. As
\onecolumngrid

\twocolumngrid

\begin{figure}[b]
\center{\includegraphics[width=1\columnwidth,angle=0,clip]{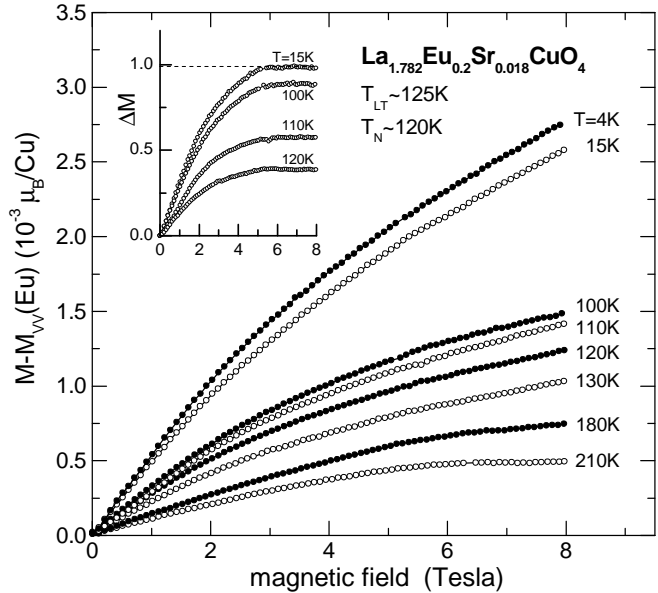}}
\caption[]{Magnetization $M(H)$ of \lesaco\ at different temperatures
after subtraction of Eu Van Vleck term $M_{VV}$(Eu). Inset: after
subtraction of linear function fitted to each curve at high fields
(6.5-8~Tesla).} \label{m_p120_pap}
\end{figure}

a consequence, for $x \gtrsim 0.017$ the \sus\ does not decrease below
$T_N$, indicating that the DM moments in adjacent layers do not arrange
AF. To study the interlayer coupling in Sr-doped \leco\ in more detail, we
have performed $M(H)$ measurements on samples with $x=0.01$ and $x=0.018$,
both annealed at 625$^\circ$C in $\rm N_2$. For $x=0.01$ we observe
basically the same behavior as for $x=0$; i.e., spin-flips for $T_{LT} < T
< T_N$, strongly decreasing spin-flips for $T < T_{LT}$ and eventually no
spin-flips for $T\ll T_{LT}$.~\cite{HueckerDiss} More interesting is the
case of $x=0.018$ with $T_N \lesssim T_{LT}$ (see Fig.~\ref{m_p120_pap}).
Here we observe weak ferromagnetic-type $M(H)$ curves for all
temperatures. The inset shows some of the curves after subtraction of the
slope between 6.5 and 8~Tesla, assuming that at these fields the
contribution of the DM moment has saturated.~\cite{maxfield} The weak
ferromagnetic behavior is obvious, though the magnetic field scale to
align the DM moments is quite high, consistent with the absence of
spontaneous weak ferromagnetism.

\subsection{Remanent moment}
\begin{figure}[t]
\center{\includegraphics[width=1\columnwidth,angle=0,clip]{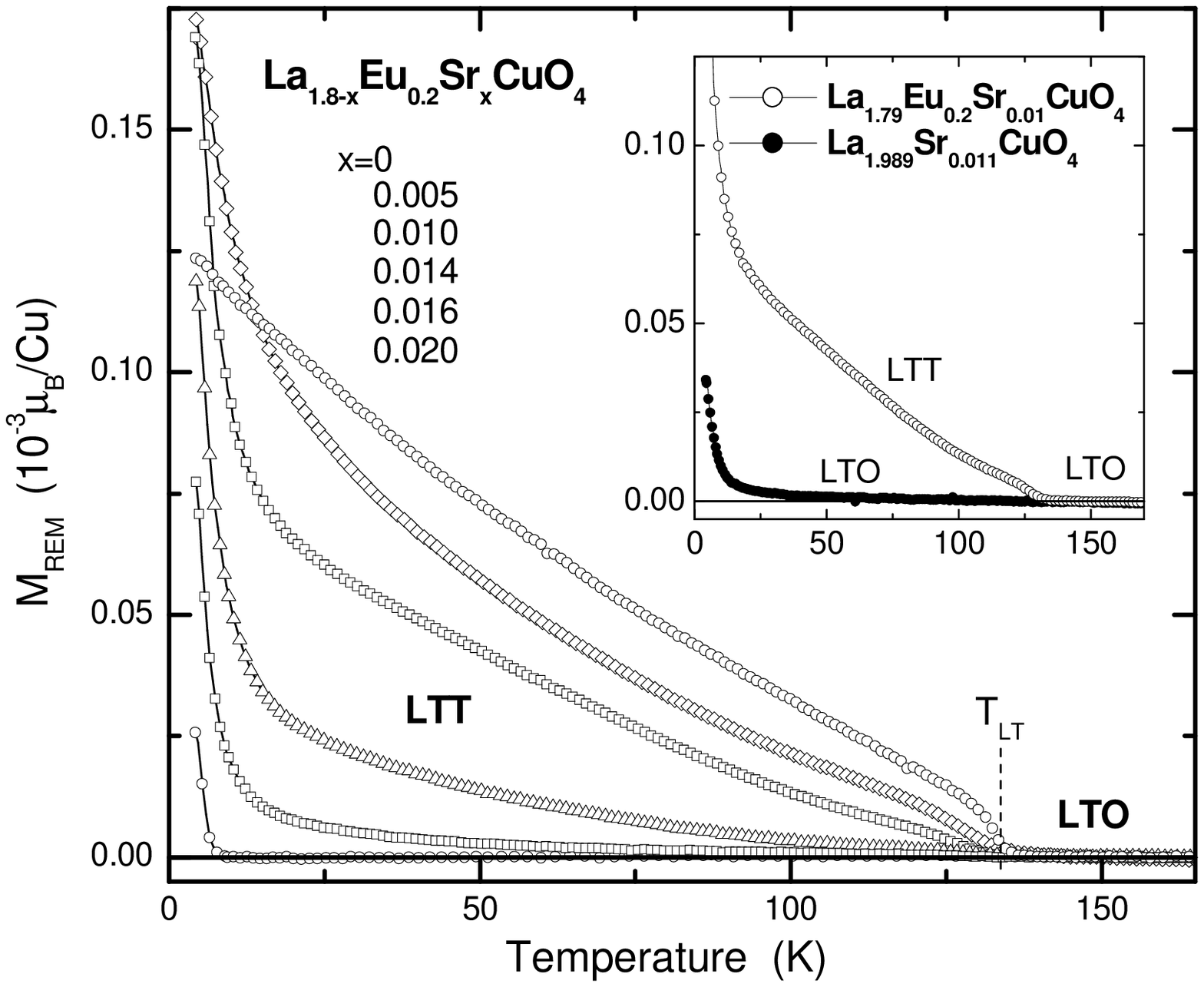}}
\caption[]{Remanent moment $M_{REM}$ of \lesco\ for different Sr
concentrations $x\leq 0.02$ as a function of temperature. Inset:
comparison with pure \lsco\ for $x\sim 0.01$.} \label{remx}
\end{figure}

In analogy to Fig.~\ref{u6_rem_pap}, the remanent moment of the Sr-doped
samples was measured in zero field after a field of 14~Tesla was applied
at 4~K. Fig.~\ref{remx} shows $M_{REM}$ for different Sr concentrations
$0\leq x \leq 0.02$ as a function of increasing temperature. With
increasing Sr content $M_{REM}$ decreases systematically and disappears at
the critical concentration $x=0.02$. $M_{REM}$ also decreases with
increasing temperature and for all $x<0.02$ vanishes at $T_{LT}$.
Obviously, a remanent moment is a feature of the entire AF ordered LTT
phase. From the comparison in the inset it is evident that in the LTO
phase, except for very low temperatures, no remanent moment is observed.
Here, $M_{REM}$ is related to the spin-freezing regime for $T\lesssim
25$~K.~\cite{Chou93,Borsa95,Niedermayer98,Matsuda02a,Suzuki02a,Huecker02b}
The spin freezing regime might also explain the upturn of $M_{REM}$ for $T
\lesssim 25$~K in the Eu-doped samples with $x>0$.~\cite{Klauss02a,Suh99b}

\begin{figure}[b]
\center{\includegraphics[width=0.98\columnwidth,angle=0,clip]{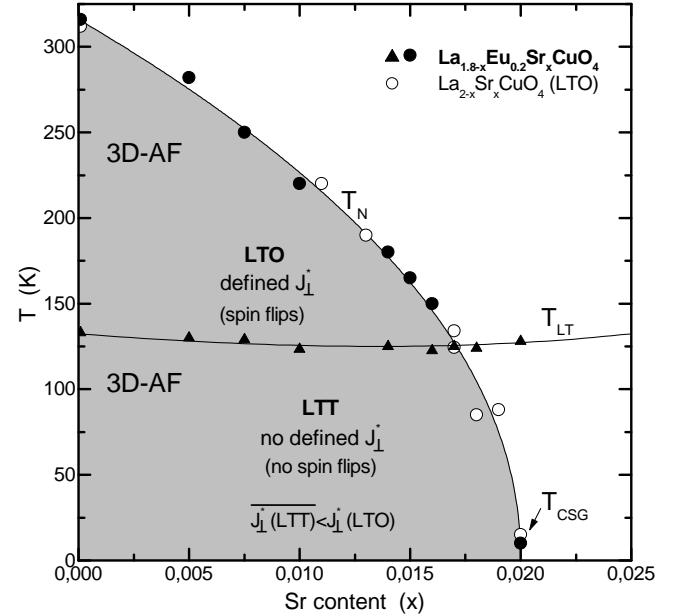}}
\caption[]{Magnetic and structural phase diagram of \lesco . Data for pure
\lco\ are shown for comparison. For $x=0.02$ the cluster spin glass
transition at $T_{CSG}$ is indicated as well. Solid lines are guides to
the eyes.} \label{EuTx_neu_pap}
\end{figure}

\subsection{Phase diagram of \lescoh}
From our measurements on \lesco\ we have constructed the phase diagram in
Fig.~\ref{EuTx_neu_pap}. For this phase diagram we have remeasured the
polycrystals after they were annealed at 800~$^\circ$C in vacuum, which
led to slightly higher N\'eel temperatures than in
Fig.~\ref{c_smallx2_pap}. In the LTO phase, the Sr doping dependence of
$T_N$ is practically identical for pure and Eu-doped \lsco . For $x >
0.016$ no N\'eel peaks are observed in the LTT phase. However, $\rm \mu$SR
experiments have shown that in the LTT phase AF order disappears at
roughly the same critical Sr content $x \simeq 0.02$ as in \lsco
.~\cite{Klauss02a} The most important difference between the LTO and LTT
phase is the loss of a well defined interlayer coupling $J_{\perp}^*$. The
weakening of the effective interlayer coupling progresses with decreasing
temperature and increasing Sr content. Accordingly, well defined spin-flip
transitions are observed only in the LTO phase, while in the LTT phase the
spin-flip disappears. Moreover, in the entire AF ordered LTT phase a small
fraction of the \DM\ spin canting can be remanently magnetized,
underlining the presence of DM moments as well as their weak ferromagnetic
character.

\section{Discussion}
\label{discussion}

As mentioned in the introduction, the question of whether in the LTT phase
Cu spins exhibit \DM\ spin canting has been a matter of
controversy.~\cite{Shamoto92,Keimer93,Yildirim95,
Koshibae94a,Stein96,Tsukada03a} Our data unambiguously show the presence
of DM spin canting in the LTT phase, supporting a spin structure given by
the black spins in Fig.~\ref{spinstructure1}. Although this is in
qualitative agreement with neutron diffraction data, there is a
significant difference with respect to the temperature dependence of the
size of the DM moments.~\cite{Keimer93}
According to Keimer et al.~\cite{Keimer93} the in-plane gap in
deoxygenated \lco\ ($T_N=325$~K) and \keimer\ ($T_N=316$~K) is about the
same in the LTO phase at 100~K, but increases by about 60\% in the LTLO
phase of the Nd doped compound. (With $\alpha\simeq 40^\circ$ the sample
is very close to LTT.) Keimer et al. argue that this increase is due to an
increase of the DM interaction in the LTLO phase. In contrast, our data
indicate a stronger DM interaction in both phases, LTO and LTT, which we
have attributed to the larger octahedral tilt angle (cf. Sec.~\ref{SC1}).

Although the increase of the in-plane gap in the LTT phase is consistent
at first glance with a larger DM moment in \leco , it is inconsistent with
the decrease of the spin-flop field $H_{c1}$ that we have observed in the
LTLO phase of the single crystal with $T_N=280$~K. Since $H_{c1}$ is a
measure of the in-plane gap, it actually indicates a decrease of this gap
in the LTLO phase, which is just the reverse of the neutron diffraction
result.~\cite{Thio90,Keimer93} Further experiments are necessary to
clarify this point. We mention that Keimer et al.~\cite{Keimer93} have
determined the anisotropy $\alpha_{DM}$ due to DM superexchange from the
in-plane gap, which is justified if the DM superexchange is by far the
dominant source. Theoretical papers, however, have pointed out that the
in-plane anisotropy may be composed from multiple finite contributions
which may compete, in particular, in the LTT
phase.~\cite{Shekhtman92,Shekhtman93,Bonesteel93b,Viertio94a,Stein96}
Tetragonal \scoc\ with flat $\rm CuO_2$ planes, for example, has the same
collinear spin structure as \lco\ but no spin canting (no DM interaction),
which shows that in this system the spin direction is determined by other
contributions to the in-plane anisotropy.~\cite{Yildirim94b} Though we do
not oppose the particular idea of a symmetric superexchange anisotropy,
our evidence of DM spin canting in the LTT phase raises serious concerns
about its quantitative relevance (cf. Sec.~\ref{background}).
Nevertheless, it is certainly a reasonable attempt to discuss the decrease
of $H_{c1}$ (spin-flop) in terms of a competition between the (enhanced)
DM interaction, which prefers a configuration with spins perpendicular to
the octahedral tilt axis, and other contributions which support other spin
directions, i.e. parallel to the tilt axis or parallel to the $b$ or
$a$-axis (cf. Fig.~\ref{spinstructure1}).

When we compare our results for the \leco\ single crystal ($T_N=315$~K)
with the data in Ref.~\onlinecite{Tsukada03a} for a less deoxygenated
crystal with $T_{N}=265$~K, we find that for $H\parallel c$ the jump at
$T_{LT}$ in our crystal, as well as the N\'eel peak, is about twice as
large, showing the strong impact of intercalated oxygen. When analyzing
our data, e.g. in Fig.~\ref{c_u6ch_pap}, in a similar way as in
Ref.~\onlinecite{Tsukada03a} [i.e., subtracting the \sus\ $\chi^c(H=8T)$
from $\chi^c(H=1T)$] we find that the difference in the LTT phase is
basically zero. While this means that in the LTT phase $M^c(H)$ is
approximately linear up to 8~Tesla, as can be seen in
Fig.~\ref{m_euu6c_pap2}, it definitely does not proof a reduction or the
absence of DM spin canting, contrary to the argument in
Ref.~\onlinecite{Tsukada03a}.

Let us assume for a moment that DM spin canting is absent. Then the fact
that in the LTT phase the effective interlayer coupling is reduced, would
bring this system very close to a perfect $S=1/2$ 2D-HAF. In this case the
\sus\ of the \pla\ planes in \leco\ should be very close to \HAF\ of
non-interacting \pla\ planes, as is the case for \lco\ in the high
temperature tetragonal phase or in \scoc .~\cite{CUaniso} Since \HAF\ is
smaller than the \sus\ of the \pla\ planes in the LTO phase, one would
expect that $\chi$ decreases at the transition
LTO$\rightarrow$LTT.~\cite{CUaniso} This is in sharp contrast to the fact
that $\chi$ actually increases (cf. Fig.~\ref{eu02eu00_pap}). Clearly, in
the absence of DM spin canting, a decrease of the interlayer coupling
cannot account for an increase of $\chi^c$ of the size observed in \lesco
, as was suggested in Ref.~\onlinecite{Tsukada03a}. Hence, in the LTT
phase Cu spins must be canted.

It is generally assumed that the interlayer coupling in the LTO phase of
\lco\ benefits from the orthorhombic strain, since the perfect frustration
of the interlayer superexchange in a tetragonal body-centered structure is
lifted.~\cite{Vaknin90,Xue88,Yildirim94,Tsukada03a} It is believed that
this is the reason for $T_N$ in \lco\ being higher than in \scoc , though
one has to keep in mind that in \scoc\ the distance between the \pla\
planes is significantly larger.~\cite{Vaknin90}
Now, in \leco\ the orthorhombic strain around $T_N$ is even bigger than in
\lco , and indeed we find a stronger interlayer coupling [cf.
Fig.~\ref{euhcneu_pap} (c)]. A significant change of $T_N$, however, is
not observed. We assume that the interlayer coupling has to change by
orders of magnitude to cause a substantial shift of $T_N$. This is also
consistent with considerations about what drives the magnetic transition
in 2D-H and 2D-XY systems in Ref.~\onlinecite{Keimer92a,Keimer92b,
Matsuda90,Ding90,Ding92,Suh95}.

In the LTT phase, a reduction of the interlayer coupling is expected
because of the tetragonal symmetry. The non-collinear spin structure is
consistent with a further weakening of the interlayer coupling (cf.
Fig.~\ref{spinstructure1}).~\cite{Thio90,Keimer93} In such a system, the
next-nearest-layer coupling might be crucial to establish a static AF
order and might result in two loosely coupled subsystems of \pla\
planes.~\cite{Keimer93} In the case of perfect decoupling of adjacent
layers, we would expect spontaneous weak ferromagnetism of the DM moments
which we have not observed. On the other hand, even if the magnetic
decoupling in the LTLO and LTT phases is not perfect, the spin lattice
should be less rigid than in the LTO phase, leading to an increase of
magnetic fluctuations. Indications for stronger fluctuations below
$T_{LT}$ have indeed been found in ESR~\cite{Kataev99b},
NQR~\cite{Suh98b,Suh99b} and $\rm \mu$SR~\cite{Klauss02a} relaxation
experiments on \lesco\ polycrystals ($0\leq x \leq 0.02$). Furthermore,
measurements of the internal static magnetic field by means of $\rm \mu$SR
indicate a slight decrease of $H_{int}^{\mu SR}$ at
$T_{LT}$.~\cite{Klauss02a} Note, however, that a change of $H_{int}^{\mu
SR}$ at the muon site can also result from slightly different muon
positions in the LTO and LTT phases. NQR measurements show a quadrupolar
broadening of the NQR-lines in the LTO phase and an even much stronger
magnetic broadening in the LTT phase. The broadening in the LTO phase was
explained with a distribution of the electric field gradient due to the
local lattice distortions caused by Eu doping.~\cite{Suh98b,Suh99b} These
lattice distortions might be responsible for the field hysteresis in the
$M(H)$ curves (cf. Fig.~\ref{m_AH88_deriv_pap}). The strong magnetic
broadening in the LTT phase was explained with a distribution of the
internal magnetic field $H_{int}^{NQR}$ at the La
site.~\cite{Suh98b,Suh99b} Since spin structure and octahedral tilts are
coupled, the magnetic broadening might indicate a distribution of the
angle $\alpha$ of the azimuthal rotation of the tilt axis. This
interpretation is consistent with our conclusion from magnetization data
that, in the LTT phase, the spin-flip field $H_c$, and therefore the
effective interlayer coupling $J_\perp^*$, are not well defined.\\

\section{Conclusion}
\label{conclusion}

In summary, we have studied the magnetism of the \pla\ planes in the
different structural phases of \lescoxy\ over a broad range of Eu and Sr
doping, with focus put on the antiferromagnetic regime. To separate the Cu
spin magnetism, we have carefully subtracted the much larger Van Vleck
magnetism of the $\rm Eu^{3+}$ ions. Our results show that \DM\
superexchange stabilizes a canted Cu spin structure in the LTT phase,
which is in sharp contradiction to theoretical predictions. Although our
result agrees with neutron diffraction data, there is disagreement with
the latter on other questions. Most intriguing is the decrease of the
spin-flop field, which suggests that the in-plane spin-wave gap decreases
in the LTLO phase while neutron scattering data indicate an increase.
Next, according to our data, the size of the canted moment of \leco ,
compared to that of pure \lco , is about 50\% larger in the LTT
\textit{and} the LTO phases, which we attribute to the larger octahedral
tilt angle. Moreover, no significant change of the canted moment was
detected at the LTO$\leftrightarrow$LTT transition itself. The major
difference of the LTT phase compared to the LTO phase is the loss of a
well defined interlayer coupling, which macroscopically results in the
disappearance of the spin-flip transition. Though the interlayer coupling
still puzzles us, it seems that it is not uniform, but has an average much
weaker than in the LTO phase. In the LTT phase, magnetization curves
become weak ferromagnetic, and exhibit a small remanent moment
perpendicular to the \pla\ planes. Spontaneous weak ferromagnetism is not
observed. The remanent moment, as well as the weak ferromagnetism, exists
only in the antiferromagentic LTT phase and disappears for $x > 0.02$
within the resolution of our experiment.

\begin{acknowledgments}

The authors thank H.-H.~Klauss, N.~Curro, P. C.~Hammel and S.~Wegner for
invaluable discussions. The work at Brookhaven was supported by the Office
of Science, US Department of Energy under Contract No. DE-AC02-98CH10886.
The work at the Universit\"at zu K\"oln am Rhein was supported by the
Deutsche Forschungsgemeinschaft through SFB~341 and SFB 608.

\end{acknowledgments}


\begin{thebibliography}{10}

\bibitem{Axe89}
J.~D.\ Axe, A.~H.\ Moudden, D.\ Hohlwein, D.~E.\ Cox, K.~M.\ Mohanty,
A.~R.\ Moodenbaugh and Y.\ Xu,
\newblock { Phys.\ Rev.\ Lett.} {\bf 62}, 2751 (1989).

\bibitem{Crawford91}
M.~K.\ Crawford, R.~L.\ Harlow, E.~M.\ McCarron, W.~E.\ Farneth, J.~D.\
Axe, H.\ Chou and Q.\ Huang,
\newblock { Phys.\ Rev. B} {\bf ~44}, 7749 (1991).

\bibitem{Buechner94c}
B.\ B\"uchner, M.\ Breuer, A.\ Freimuth and A.~P.\ Kampf,
\newblock { Phys.\ Rev.\ Lett.} {\bf 73}, 1841 (1994).

\bibitem{Tranquada95a}
J.~M. Tranquada, B.~J. Sternlieb, J.~D.\ Axe, Y.\ Nakamura and S.\ Uchida,
\newblock { Nature} {\bf 375}, 561 (1995).

\bibitem{Wagener97a}
W.\ Wagener, H.-H.\ Klauss, M.\ Hillberg M. A.~C.\ de\ Melo, M.\ Birke,
F.~J.\
  Litterst, B.\ B\"uchner and H.\ Micklitz,
\newblock { Phys.\ Rev. B} {\bf 55}, R14761 (1997).

\bibitem{Klauss00a}
H.-H.\ Klauss, W.\ Wagener, M.\ Hillberg, W.\ Kopmann, H.\ Walf, F.~J.\
  Litterst, M.\ H\"ucker and B.\ B\"uchner,
\newblock { Phys.\ Rev.\ Lett.} {\bf 85}, 4590 (2000).

\bibitem{Shamoto92}
S.\ Shamoto, T.\ Kiyokura, M.\ Sato, K.\ Kakurai, Y.\ Nakamura and S.\
Uchida,
\newblock { Physica} C~{\bf 203}, 7 (1992).

\bibitem{Crawford93}
M.~K.\ Crawford, R.~L.\ Harlow, E.~M.\ McCarron, W.~E.\ Farneth, N.\
Herron,
  H.\ Chou and D.~E.\ Cox,
\newblock { Phys.\ Rev. B} {\bf 47}, 11623 (1993).

\bibitem{Keimer93}
B.\ Keimer, R.~J.\ Birgeneau, A.\ Cassanho, Y.\ Endoh, M.\ Greven, M.~A.\
  Kastner and G.\ Shirane.
\newblock { Z.\ Physik B -- condensed matter} {\bf 91}, 373 (1993).

\bibitem{Kataev99b}
V.\ Kataev, A.\ Validov, M.\ H\"ucker, H.\ Berg and B.\ B\"uchner,
\newblock { J.\ Phys. -- Condens.\ Matter} {\bf 11}, 6571 (1999).

\bibitem{Tsukada03a}
I.\ Tsukada, X.~F.\ Sun, S.\ Komiya, A.~N.\ Lavrov and Y.\ Ando,
\newblock { Phys.\ Rev. B} {\bf 67}, 224401 (2003).

\bibitem{Suh98b}
B.~J.\ Suh, P.~C.\ Hammel, J.~L.\ Sarrao, J.~D.\ Thompson, Z.\ Fisk, M.\
H\"ucker and B.\ B\"uchner,
\newblock in {\em Physical Phenomena at High Magnetic Field III},
edited by Zachary Fisk, Lev Gor'kov and Robert Schrieffer (World
Scientific, 1999).

\bibitem{Suh99b}
B.~J.\ Suh, P.~C.\ Hammel, M.\ H\"ucker and B.\ B\"uchner,
\newblock { Phys.\ Rev. B} {\bf 59}, R3952 (1999).

\bibitem{Bonesteel93b}
N.~E.\ Bonesteel,
\newblock { Phys.\ Rev. B} {\bf 47}, 11302 (1993).

\bibitem{Shekhtman93}
L.\ Shekhtman, A.\ Aharony and O.\ Entin-Wohlman,
\newblock { Phys.\ Rev. B} {\bf 47}, 174 (1993).

\bibitem{Viertio94a}
H.~E.\ Vierti\"o and N.~E. Bonesteel,
\newblock { Phys.\ Rev. B} {\bf 49}, 6088 (1994).

\bibitem{Koshibae94a}
W.\ Koshibae, Y.\ Ohta and S.\ Maekawa,
\newblock { Phys.\ Rev. B} {\bf 50}, 3767 (1994).

\bibitem{Yildirim95}
T.\ Yildirim, A.~B.\ Harris, A.\ Aharony and O.\ Entin-Wohlman,
\newblock { Phys.\ Rev. B} {\bf 52}, 10239 (1995).

\bibitem{Stein96}
J.\ Stein, O.\ Entin-Wohlman and A.\ Aharony,
\newblock { Phys.\ Rev. B} {\bf 53}, 775 (1996).

\bibitem{Thio88}
T.\ Thio, T.~R.\ Thurston, N.~W.\ Preyer, P.~J.\ Picone, M.~A.\ Kastner,
H.~P.\
  Jenssen, D.~R.\ Gabbe, C.~Y.\ Chen, R.~J.\ Birgeneau and A.\ Aharony,
\newblock { Phys.\ Rev. B} {\bf 38}, 905 (1988).

\bibitem{Kastner88}
M.~A.\ Kastner, R.~J.\ Birgeneau, T.~R.\ Thurston, P.~J.\ Picone, H.~P.\
  Jenssen, D.~R.\ Gabbe, M.\ Sato, K.\ Fukuda, S.\ Shamoto, Y.\ Endoh, K.\
  Yamada and G.\ Shirane,
\newblock { Phys.\ Rev. B} {\bf 38}, 6636 (1988).

\bibitem{Lyons88a}
K.~B.\ Lyons, P.~A.\ Fleury, J.~P.\ Remeika, A.~S.\ Cooper and T.~J.\
Negran,
\newblock { Phys.\ Rev. B} {\bf 37}, R2353 (1988).

\bibitem{Aeppli89}
G.\ Aeppli, S.~M.\ Hayden, H.~A.\ Mook, Z.\ Fisk, S.-W.\ Cheong, D.\ Rytz,
  J.~P.\ Remeika, G.~P.\ Espinosa and A.~S.\ Cooper,
\newblock { Phys.\ Rev.\ Lett.} {\bf 62}, 2052 (1989).

\bibitem{Hayden91}
S.~M.\ Hayden, G.\ Aeppli, R.\ Osborne, A.~D.\ Taylor, T.~G.\ Perring,
S.-W.\
  Cheong and Z.\ Fisk,
\newblock { Phys.\ Rev.\ Lett.} {\bf 67}, 3622 (1991).

\bibitem{Kastner98}
M.~A.\ Kastner and R.~J.\ Birgeneau,
\newblock { Rev.\ Mod.\ Phys.} {\bf 70}, 897 (1998).

\bibitem{Thio94}
T.\ Thio and A.\ Aharony,
\newblock { Phys.\ Rev.\ Lett.} {\bf 73}, 894 (1994).

\bibitem{Moriya60}
T\^oru Moriya,
\newblock { Phys.\ Rev.\ Lett.} {\bf 4}, 228 (1960).

\bibitem{Dzyaloshinsky58}
I.\ Dzyaloshinsky,
\newblock { J.\ Phys.\ Chem.\ Solids} {\bf 4}, 241 (1958).

\bibitem{Coffey91}
D.\ Coffey, T.~M.\ Rice and F.~C.\ Zhang,
\newblock { Phys.\ Rev. B} {\bf 44}, 10112 (1991).

\bibitem{BradenDiss}
M.\ Braden,
\newblock Ph.D. thesis, Universit\"at zu K\"oln, 1992.

\bibitem{Crawford97a}
M.~K. Crawford,
\newblock in { \em High-$T_c$ Superconductivity 1996: Ten
Years after the Discovery}, edited by E.~Kaldis et~al. (Kluwer Academic
Publishers, Dordrecht, Boston, London), pp. 281-310 (1997).

\bibitem{Shekhtman92}
L.\ Shekhtman, O.\ Entin-Wohlman and A.\ Aharony,
\newblock { Phys.\ Rev.\ Lett.} {\bf 69}, 836 (1992).

\bibitem{Yildirim94b}
T.\ Yildirim, A.~B.\ Harris, O.\ Entin-Wohlman and A.\ Aharony,
\newblock { Phys.\ Rev.\ Lett.} {\bf 73}, 2919 (1994).

\bibitem{Greven95}
M.\ Greven, R.~J.\ Birgeneau, Y.\ Endoh, M.~A.\ Kastner, M.\ Matsuda and
G.\
  Shirane,
\newblock { Z.\ Physik B -- condensed matter} {\bf 96}, 465 (1995).

\bibitem{Matsuda90}
M.\ Matsuda, K.\ Yamada, K.\ Kakurai, H.\ Kadowaki, T.~R.\ Thurston, Y.\
Endoh,
  Y.\ Hidaka, R.~J.\ Birgeneau, M.~A.\ Kastner, P.~M.\ Gehring, A.~H.\ Moudden
  and G.\ Shirane,
\newblock { Phys.\ Rev. B} {\bf 42}, 10098 (1990).

\bibitem{Breuer93}
M.\ Breuer, B.\ B\"uchner, R.\ M\"uller, M.\ Cramm, O.\ Maldonado, A.\
  Freimuth, B.\ Roden, R.\ Borowski, B.\ Heymer and D.\ Wohlleben,
\newblock { Physica} C~{\bf 208}, 217 (1993).

\bibitem{Gadolinium} For the ESR experiments extra sets of polycrystals
containing 0.5\% Gd as spin probe were prepared.

\bibitem{Klauss99b}
H.-H.\ Klauss, W.\ Kopmann, D.\ Baabe, D.\ Mienert, H.\ Luetkens, F.~J.\
  Litterst, M.\ H\"ucker and B.\ B\"uchner,
\newblock {\em PSI Annual Report} (1999).

\bibitem{Klauss02a}
H.-H.\ Klauss, W.\ Wagener, W.\ Kopmann, D.\ Baabe, D.\ Mienert, F.~J.\
  Litterst, M.\ H\"ucker and B.\ B\"uchner,
\newblock { Physica B} {\bf 312}, 71 (2002).

\bibitem{Klauss03a}
H.-H.\ Klauss, D.\ Baabe, D.\ Mienert, H.\ Luetkens, F.~J.\ Litterst, B.\
  B\"uchner, M.\ H\"ucker, D.\ Andreica, U.\ Zimmermann and A.\ Amato,
\newblock { Physica B} {\bf 312}, 71 (2002).

\bibitem{Curro00a}
N.~J.\ Curro, P.~C.\ Hammel, B.~J.\ Suh, M.\ H\"ucker, B.\ B\"uchner, U.\
  Ammerahl and A.\ Revcolevschi,
\newblock { Phys.\ Rev.\ Lett.} {\bf 85}, 642 (2000).

\bibitem{Simovic03a}
B.~Simovi\u{c}, M.\ H\"ucker, P.~C.\ Hammel, B.\ B\"uchner, U.\ Ammerahl
and
  A.\ Revcolevschi,
\newblock { Phys.\ Rev. B} {\bf 67}, 224508 (2003).

\bibitem{Simovic03b}
B.~Simovi\u{c}, P.~C.\ Hammel, M.\ H\"ucker, B.\ B\"uchner and A.\
  Revcolevschi,
\newblock { Phys.\ Rev. B} {\bf 68}, 12415 (2003).

\bibitem{Kataev97a}
V.\ Kataev, B.\ Rameev, B.\ B\"uchner, M.\ H\"ucker and R.\ Borowski,
\newblock { Phys.\ Rev. B} {\bf 55}, R3394 (1997).

\bibitem{Kataev98a}
V.\ Kataev, B.\ Rameev, A.\ Validov, B.\ B\"uchner, M.\ H\"ucker and R.\
  Borowski,
\newblock { Phys.\ Rev. B} {\bf 58}, R11876 (1998).

\bibitem{JKlugeDipl}
J.\ Kluge,
\newblock Master's thesis, Universit\"at zu K\"oln, 1995.

\bibitem{Shannon76a} R.D. Shannon, Acta Chryst. A~{\bf 32}, 751 (1976)

\bibitem{Huecker97}
M.\ H\"ucker, J.\ Pommer, B.\ B\"uchner, V.\ Kataev and B.\ Rameev,
\newblock {Journal of Superconductivity} {\bf 10}, 451 (1997).

\bibitem{VanVleck}
J.~H.~Van Vleck,
\newblock {\em The Theory of Electric and Magnetic Susceptibilities}.
\newblock The International Series of Monographs on Physics. Oxford University
  Press, 1932.

\bibitem{anisotropy} The small difference between $\chi^{a^*}$ and $\chi^{b^*}$
which is visible in the LTT phase is within the experimental error, since
the crystal had to be rotated by $90^\circ$ between the two measurements.

\bibitem{Terasaki92}
I.\ Terasaki, M.\ Hase, A.\ Maeda, K.\ Uchinokura, T.\ Kimura, K.\ Kishio,
I.\
  Tanaka and H.\ Kojima,
\newblock { Physica} C~{\bf 193}, 365 (1992).

\bibitem{Tovar89}
M.\ Tovar, D.\ Rao, J.\ Barnett, S.~B.\ Oseroff, J.~D.\ Thompson, S-W.\
Cheong,
  Z.\ Fisk, D.~C.\ Vier and S.\ Schultz,
\newblock { Phys.\ Rev. B} {\bf 39}, 2661 (1989).

\bibitem{Rettori96}
C.\ Rettori, S.~B.\ Oseroff, D.\ Rao, J.~A.\ Valdivia, G.~E.\ Barberis,
G.~B.\
  Martins, J.\ Sarrao, Z.\ Fisk and M.\ Tovar,
\newblock { Phys.\ Rev. B} {\bf 54}, 1123 (1996).

\bibitem{Okabe88b}
Y.\ Okabe, M.\ Kikuchi and A.~D.~S.\ Nagi,
\newblock { Phys.\ Rev.\ Lett.} {\bf 61}, 2971 (1988).

\bibitem{Johnston97a}
D.~C.\ Johnston,
\newblock {\em Handbook of Magnetic Materials, Vol. 10}, edited by K. H.
  J.~Buschow (Elsevier Science B. V., 1997).

\bibitem{Allgeier93}
C.\ Allgeier and J.~S.\ Schilling,
\newblock { Phys.\ Rev. B} {\bf 48}, 9747 (1993).

\bibitem{Landolt}
Landolt-B\"ornstein,
\newblock volume~16 of {\em New Series, Group II}.
\newblock edited by K. H.\ Hellwege and A. M.\ Hellwege (Springer Verlag,
  Heidelberg 1986).

\bibitem{Cheong94a}
S.-W.\ Cheong, H.~Y.\ Hwang, C.~H.\ Chen, B.\ Batlogg, Jr. L.~W.\~Rupp and
  S.~A.\ Carter,
\newblock { Phys.\ Rev. B} {\bf 49}, 7088 (1994).

\bibitem{Niedermayer98}
Ch.\ Niedermayer, C.\ Bernhard, T.\ Blasius, A.\ Golnik, A.\ Moodenbaugh
and
  J.~I.\ Budnick,
\newblock { Phys.\ Rev.\ Lett.} {\bf 80}, 3843 (1998).

\bibitem{Cheong89}
S.-W.\ Cheong, J.~D.\ Thompson and Z.\ Fisk,
\newblock { Physica} C~{\bf 158}, 109 (1989).

\bibitem{Huecker02b}
M.\ H\"ucker, H.-H.\ Klauss and B.\ B\"uchner.
\newblock {cond-mat} {\bf /0210357} (2002).

\bibitem{jump} We believe that the origin for the negativ jump at $T_{LT}$ we observe
for $x=0$ and high magnetic fields is different from those negative jumps
we observe for $x>0.02$. We think that for $x=0$ the effect is connected
to the polycrystallinity of the sample, since it does not occur in the
single crystal.

\bibitem{precision} At their point of intersection, $\chi(T)$ and $M(H)$ curves
ususally agree within an error of 1-2\%. However, to increase the
precision of our analysis we have normalized the $M(H)$ curves to the
$\chi(T)$ curve at 14~Tesla.

\bibitem{Sun91a}
K.\ Sun, J.~H.\ Cho, F.~C.\ Chou, W.~C.\ Lee, L.~L.\ Miller, D.~C.\
Johnston,
  Y.\ Hidaka and T.\ Murakami,
\newblock { Phys.\ Rev.} B~{\bf 43}, 239 (1991).

\bibitem{Lavrov01a}
A.~N.\ Lavrov, Y.\ Ando, S.\ Komiya and I.\ Tsukada,
\newblock { Phys.\ Rev.\ Lett.} {\bf 87}, 017007 (2001).

\bibitem{eucontent} That the Eu content of the crystal is indeed a little
bit smaller than the nominal value of $y=0.2$ can also be seen from a
comparison of the single crystal average in Fig.~\ref{c_u6c} ($\bullet$)
and the slightly larger \sus\ of the polycrystal in
Fig.~\ref{eu_ah88_pap_combi}.

\bibitem{Thio90}
T.\ Thio, C.~Y.\ Chen, B.~S.\ Freer, D.~R.\ Gabbe, H.~P.\ Jenssen, M.~A.\
  Kastner, P.~J.\ Picone, N.~W.\ Preyer and R.~J.\ Birgeneau,
\newblock { Phys.\ Rev.} B~{\bf 41}, 231 (1990).

\bibitem{Werner00a}
R.\ Werner, M.\ H\"ucker and B.\ B\"uchner,
\newblock {Phys. Rev. B} {\bf 62}, 3704 (2000).

\bibitem{Boeni88}
P.\ B\"oni, J.~D.\ Axe, G.\ Shirane, R.~J.\ Birgeneau, D.~R.\ Gabbe,
H.~P.\
  Jenssen, M.~A.\ Kastner, C.~J.\ Peters, P.~J.\ Picone and T.~R.\ Thurston.
\newblock { Phys.\ Rev. B} {\bf 38}, 185 (1988).

\bibitem{Radaelli94}
P.~G.\ Radaelli, D.~G.\ Hinks, A.~W.\ Mitchell, B.~A.\ Hunter, J.~L.\
Wagner,
  B.\ Dabrowski, K.~G.\ Vandervoort, H.~K.\ Viswanathan and J.~D.\ Jorgensen.
\newblock { Phys.\ Rev. B} {\bf 49}, 4163 (1994).

\bibitem{Jorgensen88}
J.~D.\ Jorgensen, B.\ Dabrowski, Shiyou\ Pei, D.~G.\ Hinks, L.\ Soderholm,
B.\
  Morosin, J.~E.\ Schirber, E.~L.\ Venturini and D.~S.\ Ginley.
\newblock { Phys.\ Rev. B} {\bf 38}, 11337 (1988).


\bibitem{spinflop} The spin-flop is a first order transition only for $H\parallel b$.

\bibitem{flopfield} We mention that in a \lco\ crystal with $T_N=245$~K for
temperatures $T<100$~K a decrease of $H_{c1}$ has been observed, as
well.~\cite{Thio90} However, in this case the temperature dependence of
$H_{c1}$ is very different from that in \leco .

\bibitem{para} We mention that $M^c_{B}$ is supposed to increase with increasing
temperature upon approaching $T_N$ and for $T>T_N$ is the only nonlinear
contribution to $M^c$(H) since in the paramagnetic phase $M^c_{SF}=0$ (cf.
Ref.~\onlinecite{Thio94}).

\bibitem{strain} Due to strain related to the first-order structural
transition, the LTT phase is intrinsically inhomogeneous (see
Ref.~\onlinecite{Kataev98a}).

\bibitem{HueckerDiss}
M.\ H\"ucker,
\newblock { {}}
\newblock Ph.D. thesis, Universit\"at zu K\"oln, 2000.

\bibitem{Matsuda02a}
M.\ Matsuda, M.\ Fujita, K.\ Yamada, R.~J.\ Birgeneau, Y.\ Endoh and G.\
  Shirane,
\newblock { cond-mat} {\bf /0111228} (2001).

\bibitem{Chou93}
F.~C.\ Chou, F.\ Borsa, J.~H.\ Cho, D.~C.\ Johnston, A.\ Lascialfari,
D.~R.\
  Torgeson and J.\ Ziolo,
\newblock { Phys.\ Rev.\ Lett.} {\bf 71}, 2323 (1993).

\bibitem{Borsa95}
F.\ Borsa, P.\ Carrette, J.~H.\ Cho, F.~C.\ Chou, Q.\ Hu, D.~C.\ Johnston,
A.\
  Lascialfari, D.~R.\ Torgeson R.~J.\ Gooding, N.~M.\ Salem and K.~J.~E.\
  Vos,
\newblock { Phys.\ Rev.} B~{\bf 52}, 7334 (1995).

\bibitem{Suzuki02a}
T.\ Suzuki, T.\ Goto, K.\ Chiba, T.\ Fukase, M.\ Fujita and K.\ Yamada,
\newblock { Phys.\ Rev.} B~{\bf 66}, 172410 (2002).

\bibitem{Xue88}
W.\ Xue, G.~S.\ Grest, M.~H.\ Cohen and S.~K.\ Sinha,
\newblock { Phys.\ Rev.} B~{\bf 38}, 6868 (1988).

\bibitem{Yildirim94}
T.\ Yildirim, A.~B.\ Harris, O.\ Entin-Wohlman and A.\ Aharony,
\newblock { Phys.\ Rev.\ Lett.} {\bf 72}, 3710 (1994).

\bibitem{Vaknin90}
D.\ Vaknin, S.~K.\ Sinha, C.\ Stassis, L.~L.\ Miller and D.~C.\ Johnston,
\newblock { Phys.\ Rev.} B~{\bf 41}, 1926 (1990).

\bibitem{CUaniso} This is in particular the case for the Eu-doped polycrystals
where even small anisotropies in $\chi$ associated with the AF order of
the Cu spins in materials such as \scoc\ and \ybco\ cancel due to the
random orientation of the
crystallites.~\cite{Johnston91,Vaknin90,Tranquada88}

\bibitem{Johnston91}
D.~C.\ Johnston,
\newblock {J.\ Magnetism Magnetic Materials} {\bf 100}, 218 (1991).

\bibitem{Keimer92a}
B.\ Keimer, A.\ Aharony, A.\ Auerbach, R.~J.\ Birgeneau, A.\ Cassanho, Y.\
  Endoh, R.~W.\ Erwin, M.~A.\ Kastner and G.\ Shirane,
\newblock { Phys.\ Rev.} B~{\bf 45}, 7430 (1992).

\bibitem{Keimer92b}
B.\ Keimer, N.\ Belk, R.~J.\ Birgeneau, A.\ Cassanho, C.~Y.\ Chen, M.\
Greven,
  M.~A.\ Kastner~A.\ Aharony, Y.\ Endoh, R.~W.\ Erwin and G.\ Shirane,
\newblock { Phys.\ Rev.} B~{\bf 46}, 14034 (1992).

\bibitem{Matsuda90}
M.\ Matsuda, K.\ Yamada, K.\ Kakurai, H.\ Kadowaki, T.~R.\ Thurston, Y.\
Endoh,
  Y.\ Hidaka, R.~J.\ Birgeneau, M.~A.\ Kastner, P.~M.\ Gehring, A.~H.\ Moudden
  and G.\ Shirane,
\newblock { Phys.\ Rev.} B~{\bf 42}, 10098 (1990).

\bibitem{Ding90}
H.-Q.\ Ding and M.~S.\ Makivi\'c,
\newblock { Phys.\ Rev.\ Lett.} {\bf 64}, 1449 (1990).

\bibitem{Ding92}
H.-Q.\ Ding,
\newblock { Phys.\ Rev.\ Lett.} {\bf 68}, 1927 (1992).

\bibitem{Suh95}
B.~J.\ Suh, F.\ Borsa, L.~L.\ Miller, M.\ Corti, D.~C.\ Johnston and
D.~R.\
  Torgeson,
\newblock { Phys.\ Rev.\ Lett.} {\bf 75}, 2212 (1995).

\bibitem{Tranquada88}
J.~M.\ Tranquada, A.~H.\ Moudden, A.~I.\ Goldman, P.\ Zolliker, D.~E.\
Cox, G.\
  Shirane, S.~K.\ Sinha, D.\ Vaknin, D.~C.\ Johnston, M.~S.\ Alvarez, A.~J.\
  Jacobson, J.T.\ Lewandowski and J.~M.\ Newsam,
\newblock {Phys.\ Rev.} B~{\bf 38}, 2477 (1988).

\bibitem{maxfield} At the time of this measurement the maximum magnetic field was limited to 8~Tesla.

\end{thebibliography}
\end{document}